\documentclass[namedreferences,hyperref,optionalrh]{spr-sola}

\newcommand{\aap}{{\it Astron. Astrophys.}}

\newcommand{\apjl}{{\it Astrophys. J. Lett.}}

\chardef\us=`\_

\usepackage{amsmath}

\usepackage{natbib}
\bibpunct{(}{)}{;}{a}{}{,} 

\usepackage{graphicx}
\usepackage{txfonts}
\usepackage{siunitx}

\DeclareSIUnit[]\rsun
{\text{R\ensuremath{_{\textup{sun}}}}}

\usepackage{bm}

\usepackage[margin=1.0in]{geometry}

\begin{document} 

\begin{frontmatter}
\title{The Basic Iterative Deconvolution: A fast instrumental point-spread function deconvolution method that corrects for light that is scattered out of the field of view of a detector}

\author[addressref={aff1,aff2},email={stefan.hofmeister@columbia.edu}]{\inits{S. J.}\fnm{Stefan Johann}~\snm{Hofmeister}\orcid{0000-0001-7662-1960}}

\address[id=aff1]{Leibniz Institute for Astrophysics Potsdam, Germany}
\address[id=aff2]{Columbia University, New York, USA}

\runningauthor{S. J. Hofmeister}
\runningtitle{The Basic Iterative Deconvolution Algorithm}

\begin{abstract}
   {
   A point-spread function describes the optics of an imaging system and can be used to correct collected images for instrumental effects. The state of the art for deconvolving images with the point-spread function is the Richardson-Lucy algorithm; however, despite its high fidelity, it is slow and cannot account for light scattered out of the field of view of the detector. We reinstate the Basic Iterative Deconvolution (BID) algorithm, a deconvolution algorithm that considers photons scattered out of the field of view of the detector, and extend it for image subregion deconvolutions. Its runtime is \numrange{1.8}{7.1} faster than the Richardson-Lucy algorithm for $4096 \times 4096$ pixels images and up to an additional factor of $150$ for subregions of $250 \times 250$ pixels. We test the extended BID  algorithm for solar images taken by the Atmospheric Imaging Assembly (AIA), and find that the reconstructed intensities between BID and the Richardson-Lucy algorithm agree within $\SI{1}{\percent}$.
   }
\end{abstract}

\end{frontmatter}


\section{Introduction}

A point-spread function (PSF) describes the impulse response of an imaging system. For a perfect point source, the PSF is equivalent to the observed intensity distribution of the point source at the detector. This blurring is an effect of the imperfections of the imaging system caused by the wavelength dependence of the refractive index of optical components, slight misalignment of optical components, Fraunhofer diffraction by the pupil and metal mesh filters, and scattering of photons due to the micro-roughness of mirrors and lenses. Deconvolving an image with the instrumental PSF enables one to approximate the true image that would have been seen if the imaging system were perfect. 

The deconvolution of an image by a PSF, in general, is considered an ill-posed problem. When image deconvolution is performed by applying an inverse filter in the spatial frequency domain, the spectrum of the noise in the image becomes superimposed with the spectrum of the image, affecting the quality of the image reconstruction process. This issue can be resolved by adding a regularization to the inverse filters or by performing the deconvolution in the spatial domain. State-of-the-art in the spatial domain is the the Richardson-Lucy deconvolution algorithm \citep{richardson1972, lucy1974}, which performs the image deconvolution by an iterative Bayesian approach  based on the assumption of photon noise in the image.  

The Richardson-Lucy algorithm, though, has two disadvantages: (1) it is computationally slow and (2) it conserves flux within the image and thus does not account for flux that is scattered outside of the field of view of the detector. The slow speed impedes the PSF deconvolution of large sets of images. Not accounting for photons that have been scattered outside the field of view of the detector results in an underestimate of the intensities in the reconstructed images. This underestimate is largest near the edges of the field of view, but can even affect the center of the image if a significant fraction of light is scattered over large distances. 

These issues are of particular concern in solar physics. Satellite imagers, such as the Atmospheric Imaging Assembly (AIA; \citealt{lemen2012}) aboard the Solar Dynamics Observatory (SDO; \citealt{pesnell2012}), observe the solar corona simultaneously in multiple extreme-ultraviolet channels at a high spatial resolution and a high temporal cadence. This vast amount of data enables solar physicists to study the evolution of solar features in great detail, but also makes the image deconvolution process a tedious task. Furthermore, in the extreme-ultraviolet wavelength regime, the wavelength is at the same order of magnitude as the micro-roughness of the telescopes' mirrors, which results in diffuse long-distance scattered light \citep{deforest2009, poduval2013, auchere2023}. And the telescopes' metal mesh filters, which are used to block off-band radiation, result in a significant amount of diffracted light \citep{gburek2006, grigis2012}. Both effects result in scattering over very large image distances and entails that a portion of the photons are scattered out the field of view of the detector, impeding the image reconstruction process.

To improve on these issues, we reinstate and extend the Basic Iterative Deconvolution algorithm \citep[BID;][]{cittert1931, iinuma1967a, iinuma1967b, jansson1970a, jansson1970b}]. The BID algorithm performs the image deconvolution in the spatial domain. It neglects image noise but accounts for photons that are scattered outside of the field of view of the detector, is faster than the Richardson-Lucy algorithm, and can be extended to work on image subregions to further significantly increase its speed. At the same time, BID is flux conserving with respect to the PSF and reconstructs robustly all image regions that are not dominated by noise.

In this paper, we present the BID algorithm, extend it for the deconvolution of image subregions, and test its fidelity on AIA images of the Sun \footnote{An IDL and Python implemention of this algorithm is available at https://github.com/stefanhofmeister/PSF-Tools}. The paper is structured as follows. In Section~\ref{sec:1}, we explain basic features of PSFs and how they affect image deconvolutions to build a basis for BID and the deconvolution of image subregions. In Section~\ref{sec:2}, we present the extended BID algorithm. In Section~\ref{sec:3}, we analyze its fidelity, speed, convergence, and noise robustness, and compare its image reconstructions to the ones of the Richardson-Lucy algorithm. In Section~\ref{sec:6}, we explain how image subregions should be chosen so that subregion reconstructions are reliable. 
In Section~\ref{sec:5}, we discuss limitations of the BID algorithm and set it in the wider context of PSF deconvolutions. In Section~\ref{sec:4}, we summarize our results.


\section{The basis for BID and the deconvolution of image subregions} \label{sec:1}

\begin{figure*}
    \centering
    \includegraphics[width=\textwidth]{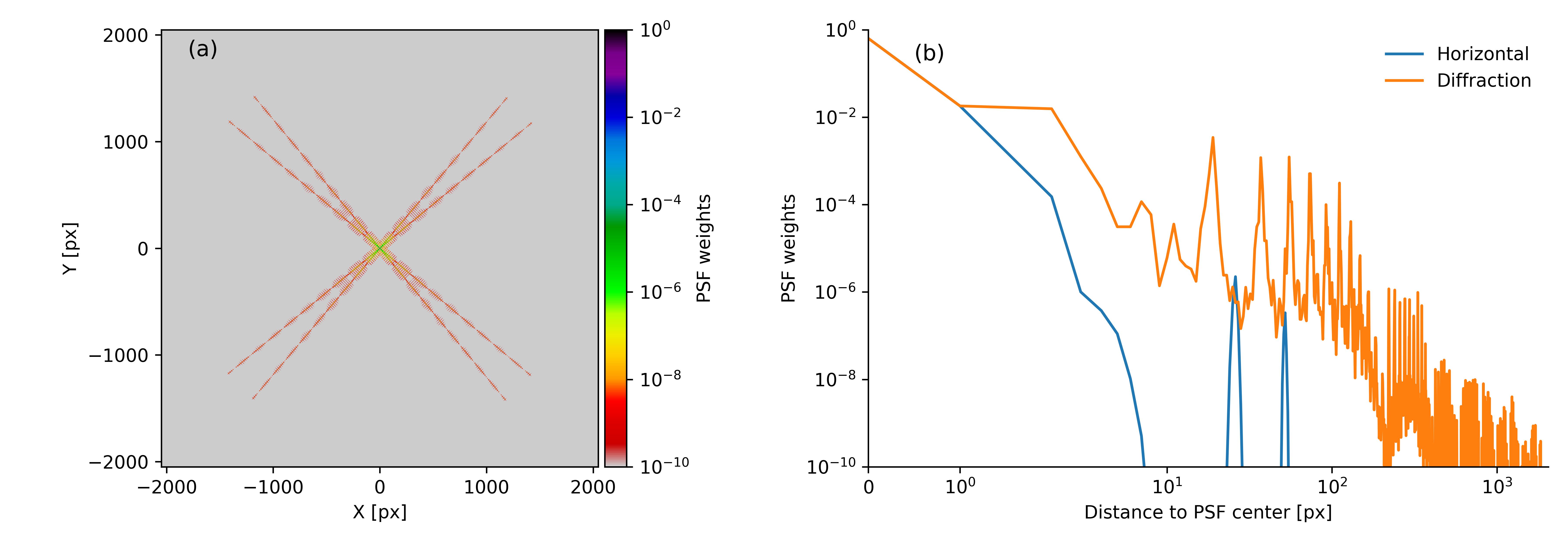}
    \caption{(a) PSF of the \SI{193}{\angstrom} channel of AIA. (b) PSF weights starting from the PSF center along a horizontal slice and along a slice through the diffraction pattern.}
    \label{fig:psf}
\end{figure*}

\begin{figure*}
    \centering
    \includegraphics[width=\textwidth]{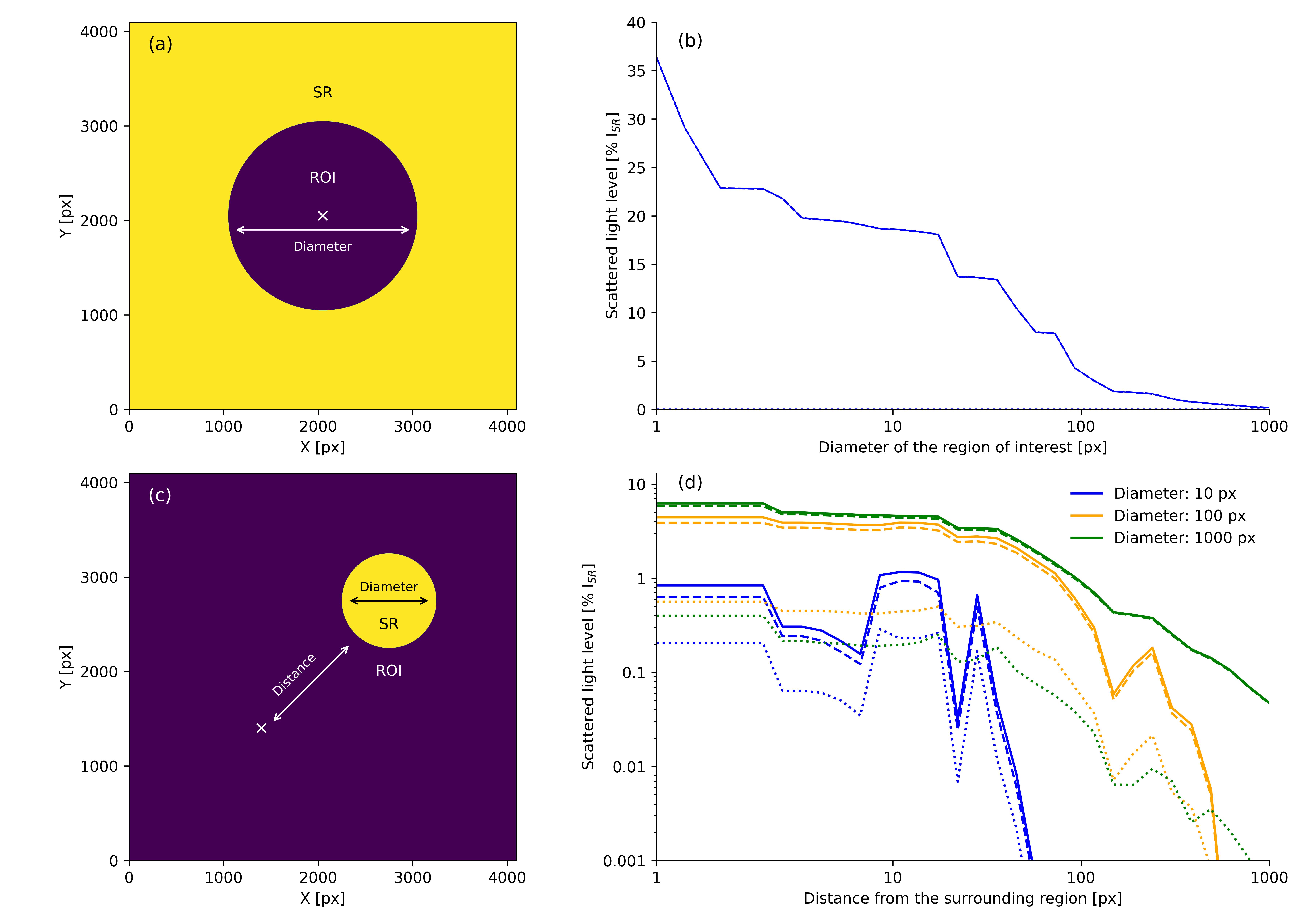}
    \caption{(a) Sketch of the first setup: we assume a homogeneous intensity in the surrounding region (SR) and vary the diameter of the region of interest (ROI). Then, we measure the level of the scattered light in the center of the ROI. (b) Scattered light level, derived with using the true intensities (solid line) and the observed intensities (dashed line). Also shown is the respective difference of the two (dotted line), which is nearly zero here. The results are shown as a function of the diameter of the ROI. (c) Sketch of the second setup. We assume the SR consists of a homogeneous intensity within a circular area, vary the diameter of the SR, and measure the level of the scattered light at a given distances to the SR. (d) Scattered light level at a given distance from the SR, derived with using the true intensities (solid lines) and the observed intensities (dashed lines). Also shown is the respective difference of the two (dotted lines). The results are shown as a function of the distance from the SR. The simulation was performed for SR diameters of $10$, $100$, and \SI{1000}{pixels}.}
    \label{fig:error_neglecting}
\end{figure*}

In this section, we first give a short overview on the properties of PSFs and the definiteness of PSF deconvolution reconstructions for PSF deconvolution algorithms that operate in the spatial domain. Then, we discuss how neglecting noise affects the image reconstruction results of spatial-domain PSF deconvolution algorithms. Finally, we study the feasibility of image subregion deconvolutions and explain how one can correct for scattered incoming light into subregion of interest and scattered outgoing light from the subregion of interest to the surrounding region. 

\subsection{The properties of PSFs and the definition of PSF deconvolution}


The PSF of an imaging system is provided in detector coordinates. The central pixel of the PSF, $\text{psf}_{\mathbf{\Delta r}=0}$ gives the amount of light that is not scattered in the instrument, and a PSF pixel located at a distance $\mathbf{\Delta r}$ from the PSF center, $\text{psf}_{\mathbf{\Delta r}}$, gives the amount of light that is scattered in direction $\mathbf{\Delta r}$ in the image plane. As no photons are lost by definition, the sum over the PSF is one.

Figure~\ref{fig:psf}(a) shows the PSF of the \SI{193}{\angstrom} channel of AIA that is provided by the instrument team, and which we use as an example throughout this study. The center coefficient of the AIA PSF is \SI{64}{\percent}, i.e., \SI{64}{\percent} of the light reaches its target destination at the detector and \SI{36}{\percent} of light undergoes scattering within the AIA instrument. The AIA PSF consists of a gaussian core and of two crosses, which are a consequence of diffraction at the entrance filter and at the focal plane filter \citep{grigis2012}. The gaussian core results in a short-distance scattering and thus a blurring of AIA images, while the diffraction arms result primarily in a directed medium-to-long-distance scattering which reduces the dynamic contrast in the images. 

The instrumental PSF degrades the quality of images observed by the instrument. 
The observed image, $\mathbf{I}_\text{o}$, is the true image, i.e., the image that would be observed by a perfect instrument, $\mathbf{I}_\text{t}$, convolved with the instrumental PSF, $\mathbf{PSF}$, plus a noise component that contains instrumental photon noise, instrumental noise, and instrumental calibration errors, $\bm{\mathcal{E}}$,
\begin{equation}
    \mathbf{I}_\text{o} = \mathbf{I}_\text{t} \ast \mathbf{PSF} + \bm{\mathcal{E}}, \label{eq:psf}
\end{equation}
where $\ast$ is the convolution operator. The aim of image deconvolution with a PSF is to reconstruct the true image from the actual observed image and the instrumental PSF. 

%
Re-expressing Eq.~\ref{eq:psf} for the observed intensity of a single pixel at a location $\mathbf{r}$ in the image plane, $I_\text{o,$\mathbf{r}$}$ gives
\begin{align}
       I_\text{o,$\mathbf{r}$} &= \sum_{\mathbf{\Delta r}} I_\text{t,$\mathbf{r}$-$\mathbf{\Delta r}$} \ \text{psf}_{\mathbf{\Delta r}} + \epsilon_{\mathbf{r}} 
       \label{eq:Io_px}
\end{align}
where the summation $\mathbf{\Delta r}$ goes over the entire PSF, $I_\text{t,$\mathbf{r}$-$\mathbf{\Delta r}$}$ is the intensity of a pixel at the location $\mathbf{r}-\mathbf{\Delta r}$, 
and $\epsilon_{\mathbf{r}}$ is the noise within the observed pixel. This summation describes both the portion of light that correctly reaches the target pixel given by $I_\text{t,$\mathbf{r}$} \ \text{psf}_{\mathbf{\Delta r}=0}$, and the light scattered   from other pixels into the target pixel given by $I_\text{t,$\mathbf{r}$-$\mathbf{\Delta r}$} \ \text{psf}_{\mathbf{\Delta r} > 0}$.
Equation~(\ref{eq:Io_px}) holds for each pixel in the image plane and thus sets up a system of equations. If one neglects the noise component, the true intensities can be approximated by a multi-linear fit. Mathematically, this system of equations and thus image deconvolutions in general are ill-posed, as the noise component and zero-amplitude components of frequencies in the PSF inhibit a unique solution. However, in practice, experience has shown that PSF deconvolutions correct the instrumental effects for large and thereby strongly enhance the quality of the collected images.

\subsection{The effect of noise} \label{subsec:noise}

The question is if one is allowed to neglect the noise component in the right-hand side of Eq.~\ref{eq:Io_px} during the image deconvolution? In the following, we study this question by deriving the involved error in the reconstructed true
intensities.
Neglecting the noise component  in the right-hand side of Eq.~\ref{eq:Io_px}  and dubbing the derived solution for the true intensities as $\mathbf{I}_\text{t,der}$ turns Eq.~\ref{eq:psf} to 
\begin{equation}
    \mathbf{I}_\text{o} = \mathbf{I}_\text{t,der} \ast \mathbf{PSF}. \label{eq:psf_noise}
\end{equation}

The derived true intensities can be written as a composition of the true intensities and the, due to the neglected noise, deviations from the true intensities, $\mathbf{I}_\text{t,dev}$,
\begin{equation}
    \mathbf{I}_\text{t,der} = \mathbf{I}_\text{t} + \mathbf{I}_\text{t,dev}. \label{eq:noise_def1}
\end{equation}
The observed intensities, $\mathbf{I}_\text{o}$, can be written as a composition of the observed intensities when no noise would be present, $\mathbf{I}_\text{o,$\epsilon$=0}$, and the present image noise, 
\begin{equation}
    \mathbf{I}_\text{o} = \mathbf{I}_\text{o,$\epsilon$=0} + \bm{\mathcal{E}}.     \label{eq:noise_def2}
\end{equation}
Using the definitions of Eq.~\ref{eq:noise_def1} and Eq.~\ref{eq:noise_def2}, Eq.~\ref{eq:psf_noise} turns to
\begin{equation}
     \mathbf{I}_\text{o,$\epsilon$=0} + \bm{\mathcal{E}} = \left( \mathbf{I}_\text{t} + \mathbf{I}_\text{t,dev}  \right) \ast \mathbf{PSF}. \label{eq:noise1} 
\end{equation}
Subtracting from Eq.~\ref{eq:noise1} the noise-free case of Eq.~\ref{eq:psf}, i.e., 
\begin{equation}
\mathbf{I}_\text{o,$\epsilon$=0} = \mathbf{I}_\text{t}  \ast \mathbf{PSF} ,
\end{equation}
finally defines the deviations of the derived true intensities from the real true intensities, 
\begin{equation}
      \bm{\mathcal{E}} = \mathbf{I}_\text{t,dev}  \ast \mathbf{PSF}. \label{eq:noise2_2}
\end{equation}
Therefore, the deviations from the true image, $\mathbf{I}_\text{t,dev}$, which corresponds to the error in the reconstructions due to the neglected noise component in the right-hand side of Eq.~\ref{eq:Io_px}, can be estimated by deconvolving the image noise maps with the PSF.

For instrumental PSFs, most of the energy of the PSF kernel  is located in the center of the PSF. For the AIA PSF, \SI{64}{\percent} of the energy is located in the center coefficient of the PSF, \SI{1.8}{\percent} in the horizontally adjacent pixels to the center, and only \SI{0.015}{\percent} in pixels being 2 pixels away from the PSF center in the horizontal direction \citep{grigis2012}. The remaining PSF weights at larger distances from the PSF center are less important, as the noise map over larger distances is on the average close to zero. As most of the energy is concentrated in the PSF core and larger distances can be disregarded, the error in the reconstructions only depend on the local noise level.

When we further approximate that the adjacent pixels to the PSF center are negligible compared to the PSF center, we can derive an analytical approximation for the error in the reconstructions. For an image pixel located at $\mathbf{r}$ in the image, Eq.~\ref{eq:noise2_2} then turns to 
\begin{equation}
     \epsilon_{\mathbf{r}} = I_\text{t,dev,$\mathbf{r}$}\  \text{psf}_{\mathbf{\Delta r}=0},
\end{equation}
and the error in the reconstructions can be roughly approximated by 
\begin{equation}
     I_\text{t,dev,$\mathbf{r}$} =  \frac{\epsilon_{\mathbf{r}}}{\text{psf}_{\mathbf{\Delta r}=0} } .
\end{equation}
For instrumental PSFs, the center coefficient of the PSF is typically in the range \numrange{0.3}{1.0}; if it is smaller, the instrument is either not well focused or has serious technical design issues. Therefore, for instrumental PSFs, the error in the reconstructions by neglecting the image noise is typically in the range of $1\text{--}3\ \epsilon_{\mathbf{r}}$ (for AIA: $1.56\ \epsilon_{\mathbf{r}}$), i.e., at the order of the noise level.

\subsection{Deconvolving image subregions} \label{subsec:subfields}

Image deconvolution can be a computationally intensive task. If we could constrain the deconvolution to image subregions, we would greatly speed up this process for many use cases. To evaluate if the deconvolution of subregions is justifiable, we need to study the error that is involved in neglecting the surrounding image regions. We show that this is most likely to be justifiable for image subregions that are much brighter than the surrounding image regions, i.e., bright regions such as solar active regions and flares.

Let us split the image into two regions: (1) the region of interest, which we label by the index $\text{ROI}$, and (2) the surrounding region, which we label by the index $\text{SR}$. Using this definition, for an observed intensity in the region of interest at a location $\mathbf{r}$, Eq.~(\ref{eq:Io_px}) becomes 
\begin{equation}
        I_\text{o,$\mathbf{r} \in \text{ROI}$} = \sum_{ (\mathbf{r}-\mathbf{\Delta r}) \ \in \ \text{ROI}} I_\text{t,$\mathbf{r}$-$\mathbf{\Delta r}$}  \ \text{psf}_{\mathbf{\Delta r}} + \sum_{ (\mathbf{r}-\mathbf{\Delta r}) \ \in \ \text{SR}} I_\text{t,$\mathbf{r}$-$\mathbf{\Delta r}$}  \ \text{psf}_{\mathbf{\Delta r}} + \epsilon_{\mathbf{r}}. \label{eq:ROI_SR}
\end{equation}

We first assume that the intensity in the pixels in the surrounding region are equal, $I_\text{t,$(\mathbf{r}$-$\mathbf{\Delta r}) \in \text{SR}$} = I_\text{SR}$. Using the property that the sum over the PSF is one and thus that a partial sum over the PSF is smaller than one, the second term of Eq.~(\ref{eq:ROI_SR}) simplifies to 
\begin{equation}
    \sum_{ (\mathbf{r}-\mathbf{\Delta r}) \ \in \ \text{SR}} I_\text{t,$\mathbf{r}$-$\mathbf{\Delta r}$}  \ \text{psf}_{\mathbf{\Delta r}} = I_\text{SR} \sum_{ (\mathbf{r}-\mathbf{\Delta r}) \ \in \ \text{SR}}  \ \text{psf}_{\mathbf{\Delta r}} < I_\text{SR}.
\end{equation}
Therefore, the error in neglecting the second term in Eq.~(\ref{eq:ROI_SR}), i.e., the incoming light scattered from the surrounding region, is smaller than the intensity level in the surrounding region. By how much depends on the PSF and on the distance between the observed pixel in the region of interest and the edge of the surrounding region. For the PSF of the \SI{193}{\angstrom} channel of AIA, we derive an upper limit for that  error by simulating the scattered light level from the surrounding region into the region of interest. We set the true intensities of the homogeneous surrounding region to a given intensity level, $I_\text{SR}$, add a circular region of interest that has an intensity of zero, convolve this setup with the PSF, and measure the scattered light level in the center of the region of interest. The setup and results are shown in Fig.~\ref{fig:error_neglecting}(a) and (b), respectively. For a region of interest with a diameter of \SI{10}{pixels}, the scattered light level in its center is $\SI{19}{\percent} \  I_\text{SR}$, and for a diameter of \SI{100}{pixels}, the scattered light level in its center is $\SI{4}{\percent} \ I_\text{SR}$.

Next, we test what error we can expect when we neglect a bright spot in the surrounding region at a given distance to a point in the region of interest. To do so, we assume that the bright spot has a circular shape and a homogeneous intensity, vary the diameter of the bright spot, and measure the scattered light level at a given distance to the bright spot. The setup and results for the \SI{193}{\angstrom} channel of AIA are shown in Fig.~\ref{fig:error_neglecting}(c) and (d), respectively. The scattered light level is between \SI{0}{\percent} and $\SI{6.2}{\percent} \ I_\text{SR}$, depending on the distance of the observed pixel to the spot, the size of the spot, and the brightness of the spot. The irregular shape of the scattered light level in Fig.~\ref{fig:error_neglecting}(d) results from the diffraction pattern of the AIA PSF. 

These results show that light scattered into the region of interest from the surrounding region is up to \SI{20}{\percent} of $ I_\text{SR}$. It can only be neglected if $I_\text{SR} \ll I_\text{o,$\mathbf{r}$}$, i.e., if the surrounding region has a much smaller intensity level than the region of interest and if there are no bright spots in the surrounding region that are close to the region of interest. For solar coronal images, these conditions are likely best fullfilled for active regions and flares. In order to deconvolve a subregion where the surrounding region has a similar or higher intensity than the subregion, an estimate on the light scattered from the surrounding region into the subregion is required.

\subsection{Estimating the light that is scattered into the region of interest from the surrounding region} \label{sec:scatteredlight}

How can we get an estimate on the light that is scattered into the region of interest from the surrounding? A simple approach is to assume that the large-scale spatial distribution of the true intensities in the surrounding region is similar to the large-scale spatial distribution of the observed intensities in the surrounding region, i.e., that 
\begin{equation}
    \sum_{ (\mathbf{r}-\mathbf{\Delta r}) \ \in \ \text{SR}} I_\text{t,$\mathbf{r}$-$\mathbf{\Delta r}$}  \ \text{psf}_{\mathbf{\Delta r}} \approx \sum_{ (\mathbf{r}-\mathbf{\Delta r}) \ \in \ \text{SR}} I_\text{o,$\mathbf{r}$-$\mathbf{\Delta r}$}  \ \text{psf}_{\mathbf{\Delta r}} .\label{eq:scattered_light_estimate}
\end{equation}
This assumption is usually justified as most of the PSF energy is contained within the PSF core, i.e., that most photons are only scattered a short distance of several pixels away from their expected location.
Analogous to Section~ \ref{subsec:subfields}, we study this assumption for a homogeneous surrounding region and for a surrounding region that contains a bright spot. 

We first assume a homogeneous image, where the true intensity in each pixel is equal. Since the sum over the PSF is one, Eq.~(\ref{eq:Io_px}) simplifies to 
\begin{equation}
    I_\text{o,$\mathbf{r}$} =  \sum_{\mathbf{\Delta r}} I_\text{t,$\mathbf{r}$-$\mathbf{\Delta r}$} \ \text{psf}_{\mathbf{\Delta r}} = I_\text{t,$\mathbf{r}$} \sum_{\mathbf{\Delta r}} \text{psf}_{\mathbf{\Delta r}} = I_\text{t,$\mathbf{r}$}.
\end{equation}
Thus, in a homogeneous image, the observed intensities equal the true intensities. Therefore, for the purpose of estimating the scattered light level arising from a homogeneous surrounding region, one might use the observed intensities in the surrounding region as a proxy for its true intensities.
We test this assumption on the setup shown in Fig.~\ref{fig:error_neglecting}(a). This time, we first convolve the true intensities in the surrounding region with the PSF to derive the observed intensities in the surrounding region. We then use these intensities in the observed surrounding region as a proxy for the intensities in the true surrounding region. Next, analogous to Section \ref{subsec:subfields}, we add a circular region of interest that has an intensity of zero and convolve this setup with the PSF to derive an estimate of the scattered light in the region of interest. The result is shown as dashed lines in Fig.~\ref{fig:error_neglecting}(b).  The simulated scattered light level using the observed intensities as proxies follows well the simulated scattered light level derived from the true intensities; the difference is only at the order of $\SI{e-5}{\percent} \ I_\text{SR}$. 

We perform a similar simulation for the setup shown in Fig.~ \ref{fig:error_neglecting}(c), i.e., which contains a bright spot in the surrounding region. We again use the observed intensities of the bright spot as proxy for its true intensities and measure the scattered light level, depending on distance from an arbitrary point in the region of interest to the bright spot and on the size of the bright spot. The result is shown as dashed lines in Fig.~\ref{fig:error_neglecting}(d). The simulated scattered light using the observed intensities as proxy slightly underestimates the true scattered light level; the error is between of $\SI{0.6}{\percent} \  I_\text{SR}$  and $\SI{0.2}{\percent} \  I_\text{SR}$ for distances smaller than \SI{12}{pixels}, and falls below $\SI{0.1}{\percent} \ I_\text{SR}$ for distances larger than \SI{100}{pixels}. 

These estimates show that Eq.~(\ref{eq:scattered_light_estimate}) provides an acceptable proxy for the scattered light as long as the surrounding region can be approximated as homogeneous or as long as bright spots are not very close to the region of interest. For solar physics, this condition is well met for coronal quiet Sun regions as long as an active region is not very close to the region of interest. 

\subsection{Accounting for light scattered from the region of interest into the surrounding region} \label{sec:conservationofflux}
In the previous section, we have discussed how one can estimate the incoming light scattered from the surrounding region into the region of interest. How can we account for the outgoing light scattered from the region of interest into the surrounding region? This question is related to the conservation of flux.

The PSF is flux conserving, i.e., the sum over the PSF is one, and as such, no photons are lost nor generated in the image plane. However, the region of interest generally only covers a part of the image plane. Let us assume that all intensity sources are within the region of interest. Due to the scattering, some photons will be scattered out of the region of interest, and, consequentially, the total observed flux in the region of interest is smaller than the total true flux in the region of interest. If we use a deconvolution method that conserves flux within the region of interest, such as the Richardson-Lucy algorithm, the photons that are scattered out of the region of interest cannot be retrieved. The true flux in the region of interest will be underestimated. Therefore, one should not require that in the region of interest the flux is conserved during the PSF deconvolution. Instead,  one should require that the solution for the true image follows the governing equation that describes the scattering by the PSF, i.e., Eq.~(\ref{eq:psf}), or, equivalently, Eq.~(\ref{eq:Io_px}) and Eq.~(\ref{eq:ROI_SR}), and one should rely on the feature that the PSF itself is flux conserving when considering the entire image plane. 

Is it sufficient to solve Eq.~(\ref{eq:ROI_SR}) for the region of interest to account for light scattered out of the region of interest? Let us assume that the region of interest has a good signal-to-noise ratio, i.e., that we can neglect the noise component. Further, we use the observed intensities in the surrounding region as a proxy for its true intensities, i.e.,  Eq.~(\ref{eq:scattered_light_estimate}), to estimate the incoming light scattered from the surrounding into the region of interest. Then, Eq.~(\ref{eq:ROI_SR}) becomes
\begin{equation}
     \sum_{ (\mathbf{r}-\mathbf{\Delta r}) \ \in \ \text{ROI}} I_\text{t,$\mathbf{r}$-$\mathbf{\Delta r}$}  \ \text{psf}_{\mathbf{\Delta r}}\ =\ I_\text{o,$\mathbf{r} \in \ \text{ROI}$}\ -\ \sum_{ (\mathbf{r}-\mathbf{\Delta r}) \ \in \ \text{SR}} I_\text{o,$\mathbf{r}$-$\mathbf{\Delta r}$}  \ \text{psf}_{\mathbf{\Delta r}}  \label{eq:fluxcons}
\end{equation}
Equation~\ref{eq:fluxcons} has to hold for each pixel in the region of interest and thus sets up a system of equations. The observed intensities in the System of Equations~\ref{eq:fluxcons} are known, the PSF coefficients are known, and we have to fit one true intensity per pixel in the region of interest, i.e., we have to fit as many variables as we have    lines in the system of equations. Thus, System of Equations~\ref{eq:fluxcons} is well-defined,  and it follows that its solution for the true intensities is unique. The solution depends only on the PSF, on the intensity distribution in the region of interest, and on the incoming scattered light from the surrounding region. It does not explicitly depend on the outgoing scattered light from the region of interest to the surrounding region; the information on the outgoing scattered light is implicitly included in the PSF. Since we started our derivation with the governing Equation on PSFs, i.e., Eq.~(\ref{eq:psf}), and since we have nowhere neglected the outgoing scattered light during our derivation, it follows that Eq.~(\ref{eq:fluxcons}) accounts for photons that are scattered out of the region of interest. 

In the following, we demonstrate that Eq.~(\ref{eq:fluxcons}) accounts for photons that are scattered out of the region of interest on the example of three types of regions of interest in solar images. 
First, we assume that the region of interest is a single flaring pixel, i.e., that the true intensity in the flaring pixel is much larger than the intensities in all other pixels. In this case, the noise in the flaring pixel and the incoming light scattered from all other pixels can be neglected, and  Eq.~(\ref{eq:fluxcons}) becomes 
\begin{equation}
    I_\text{t,$\mathbf{r}$} = \frac{I_\text{o,$\mathbf{r}$}}{\text{psf}_{\mathbf{\Delta r}=0} }.
\end{equation}
Thus, the true intensity of the flaring pixel is reconstructed by using only the information of the observed intensity of the flaring pixel and the central coefficient of the PSF. The recorded intensity distribution of the scattered light in the surrounding of the flaring pixel is not required, as this information is implicitly included in the PSF.

Second, the region of interest is an isolated active region within a quiet Sun environment, i.e., an extended object that has a much higher intensity than its surrounding. Again, the noise component and the incoming light scattered from the surrounding can be neglected. For a single pixel in the active region, Eq.~(\ref{eq:fluxcons}) becomes 
\begin{equation} 
    I_\text{o,$\mathbf{r}$} = \sum_{ (\mathbf{r}-\mathbf{\Delta r}) \ \in \ \text{ROI}} I_\text{t,$\mathbf{r}$-$\mathbf{\Delta r}$}  \ \text{psf}_{\mathbf{\Delta r}}\label{eq:Io_extended}
\end{equation}
Therefore, the observed intensity in the active region only depends on the true intensities in the active region and on the PSF coefficients. As the PSF is known and as we have as many observed intensities in the active region as true intensities to determine  (one observed intensity and one true intensity per active region pixel), Eq.~(\ref{eq:Io_extended}) is a well-defined system of equations that can be solved for the true active region intensities. As this system of equations is well-defined, its solution is unique. No knowledge on the observed intensities in the region surrounding of the active region is required to reconstruct the true active region intensities.

And third, we assume that an entire solar image including its off-limb regions is the region of interest, and that the PSF has a long tail, i.e., that a significant amount of photons are scattered outside the field of view of the detector. As the intensities in interplanetary space outside the detector are negligible compared to the solar intensities, the incoming light scattered from interplanetary space that is outside of the detector can be neglected. We again end up at Eq.~(\ref{eq:Io_extended}), whose solution is unique.
Therefore, we can correctly reconstruct the true image intensities for the entire solar image, even when the PSF scatters a significant amount of photons outside the field of view of the detector.

\section{The extended BID algorithm} \label{sec:2}
In Section~\ref{sec:1}, we have shown several features of instrumental spatial-domain PSF deconvolutions: (1) that the error in the reconstructed true intensities by neglecting noise is at the order of the local noise level, (2) that the information on the outgoing scattered light from the subregion of interest to the surrounding is implicitly included in the PSF, and (3) that a subregion deconvolution is possible if the incoming scattered light level from surrounding region to the subregion of interest is negligible or can be approximated from the surrounding region. In this section, we present the actual algorithm, BID.

The governing equation for PSF deconvolutions is Eq.~(\ref{eq:psf}). BID aims at reconstructing the true image under the assumption that the image noise is negligible and under the constraint that the derived true image convolved with the PSF gives the observed image. Therefore, BID solves the equation 
\begin{equation}
        \mathbf{I}_\text{o} = \mathbf{I}_\text{t} \ast \mathbf{PSF}. \label{eq:bid}
\end{equation} 
by minimizing the difference between an approximated true image convolved with the PSF and the observed image,
\begin{equation}
   \text{min} \left( \lVert  \mathbf{I}_\text{t} \ast \mathbf{PSF} - \mathbf{I}_\text{o}  \rVert \right). \label{eq:minimization}
\end{equation}
To determine the solution for the true intensities from the System of Equations~\ref{eq:bid}, BID uses the following standard iterative solver in image processing. This solver corresponds mathematically to the Jacobi method for solving a multilinear system of equations in signal processing applications. For details on the solver, we refer to \citet{skarsgard1961}. We extend the BID algorithm by two preprocessing steps which enable the desired subregion deconvolution:

\begin{enumerate}
    \item In case of subregion devonvolution: cut out the subregion of interest.
    \item In case of subregion deconvolution: estimate the scattered light level from the surrounding region into the subregion of interest and subtract it from the observed image.
    \item BID: Set the observed image as the approximated true image, $$\mathbf{I}_\text{o} \Rightarrow  \mathbf{I}_\text{t}.$$
    \item \label{enum:start_iterate} BID: Convolve the approximated true image with the PSF to derive a simulated observed image, $$ \mathbf{I}_\text{t} \ast \mathbf{PSF} \Rightarrow \mathbf{I}_\text{o,sim}. $$
    \item BID: Derive pixelwise the deviation between the simulated observed image and the observed image, $$ \mathbf{I}_\text{o,sim} - \mathbf{I}_\text{o} \Rightarrow \mathbf{I}_\text{dev}. $$
    \item BID: Subtract this deviation from the approximated true image, $$  \mathbf{I}_\text{t}  - \mathbf{I}_\text{dev} \Rightarrow \mathbf{I}_\text{t}.  $$
    \item BID: Repeat from (\ref{enum:start_iterate}) until this deviation becomes negligible, i.e., until BID algorithm has converged.
\end{enumerate}


The quality of convergence, speed, and accuracy of the algorithm for AIA images will be investigated in the next section.
In the following, we  describe the required steps for an implementation of the algorithm in more detail. \\[-.2cm] 

\noindent\textbf{1. Cutting out the region of interest}
In case that a subregion deconvolution is desired, we first cutout the region of interest from the image. Since the PSF has the same dimensions as the original image, we also have to cut out a PSF subfield to adjust the size of the PSF accordingly. Since scattering is described by the PSF weights with respect to the PSF center, (1) the center of the cutout PSF has to coincide with the center of the original PSF, and (2)  the dimensions of the cutout PSF must be twice the dimensions of the region of interest so that the cutout PSF describes scattering over the entire region of interest in all directions.
However, for computational efficiency, the dimensions of the PSF and of the image should be equal during the deconvolution procedure. Therefore, we pad the cutout region of interest with zeros to reach the dimensions of the cutout PSF. \\[-.2cm] 

\noindent\textbf{2. Estimating the light scattered from the surrounding region into the region of interest}
We first set the intensities of the region of interest in the original image to zero and then convolve this image with the PSF. The intensities in the region of interest of the resulting image resemble the incoming light scattered from the surrounding region into the region of interest. We subtract the incoming scattered light from the observed intensities.\\[-.2cm] 

\noindent\textbf{3. First approximation for the true image}
Basically, one can use an arbitrary choice for the first approximation for the true image. Since we know that the true intensities are close to the observed intensities for sufficiently large homogeneous image regions (see Sect.~\ref{sec:scatteredlight}), the observed image yields a preferable choice for the first approximation of the true image.\\[-.2cm] 

\noindent\textbf{4. Simulating an observed image}
To create a simulated observed image, we convolve the approximated true image with the PSF. For computational efficiency, we recommend performing the convolution operation in the spatial frequency domain. The convolution operation in the spatial frequency domain yields periodic boundary conditions, i.e., the light that is scattered outside of the image at one edge comes in again at the opposing edge. To break the periodic boundary conditions, the PSF and the image have to be padded with zeros equal to the width of half of their dimensions before their are transformed to the spatial frequency domain.\\[-.2cm] 

\noindent\textbf{5. and 6. Deriving a revised approximation for the true image}
To derive a revised approximation for the true image, we derive pixelwise the difference between the simulated observed image and the observed image, and subtract this difference from the approximated true image.\\[-.2cm] 

\noindent\textbf{7. Check if the algorithm has converged} 
The algorithm has converged when the deviation between the observed image and the simulated observed image becomes smaller than a given metric. We define the metric to be the maximum deviation between the observed image and the simulated observed image, and the termination condition as the maximum deviation between the observed image and simulated observed image becoming smaller than $0.1$ digital numbers (DNs; \SI{1}{DN} corresponds to one intensity count in the image).
Until the algorithm has converged, we repeat the procedure from Step 4.

\section{Test cases} \label{sec:3}

\begin{figure*}
    \centering
    \includegraphics[width=.7\textwidth]{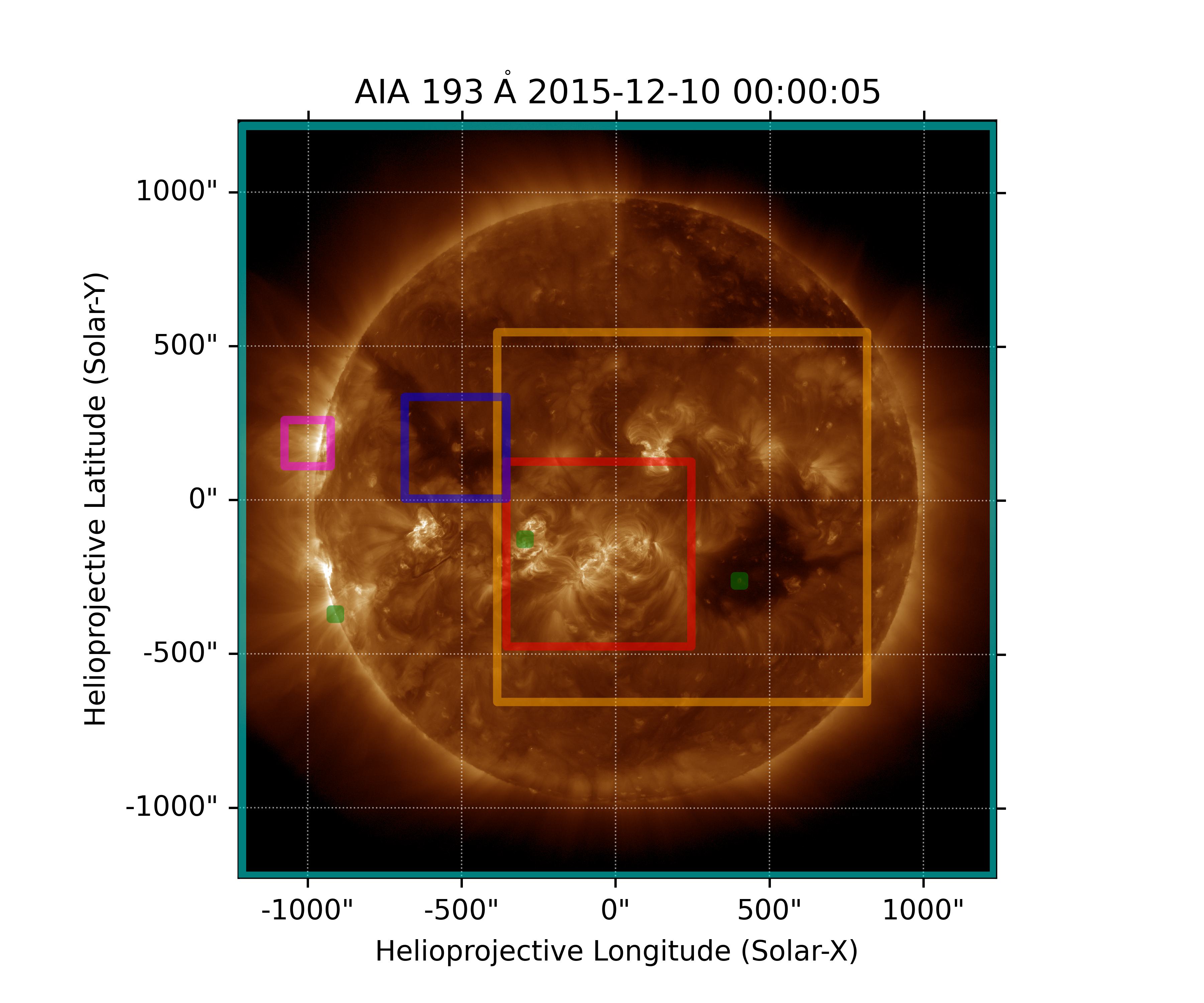}
    \caption{AIA-\SI{193}{\angstrom} image, taken on 10 December 2015. The colored boxes mark selected subregions.}
    \label{fig:overview}
\end{figure*}

Having set up the algorithm, we analyze its performance by comparing its results to the results of the Richardson-Lucy deconvolution algorithm. 
We set the termination condition of BID to \SI{0.1}{DN}, and set the Richardson-Lucy algorithm run for a fixed number of $25$~iterations, which is the standard setting for guaranteeing a good convergence in solar physics.
We take the AIA-\SI{193}{\angstrom} image recorded on 10 December 2015 with a resolution of $4096\times 4096$~pixels as test case (Fig.~ \ref{fig:overview}) and study the fidelity, convergence, speed, and noise robustness of BID by deconvolving the entire image as well as the seven marked subregions with the AIA PSF. In addition, we study the accuracy of BID for a statistical set of AIA images that covers the time range from 2010 to 2023.  

We perform these tests on a NVIDIA~GeForce~RTX~2080 graphics processing unit (GPU) and on an AMD Ryzen~9 3950X central processing unit (CPU), which are consumer-level computing components, for a Python GPU, a Python CPU, and an IDL CPU implementation of the algorithms. Unless otherwise stated, the following results were produced using the Python GPU implementation. 

\subsection{Proof of Concept} \label{subsec:proof_of_concept}

\begin{figure*}
    \centering
    \includegraphics[width=\textwidth]{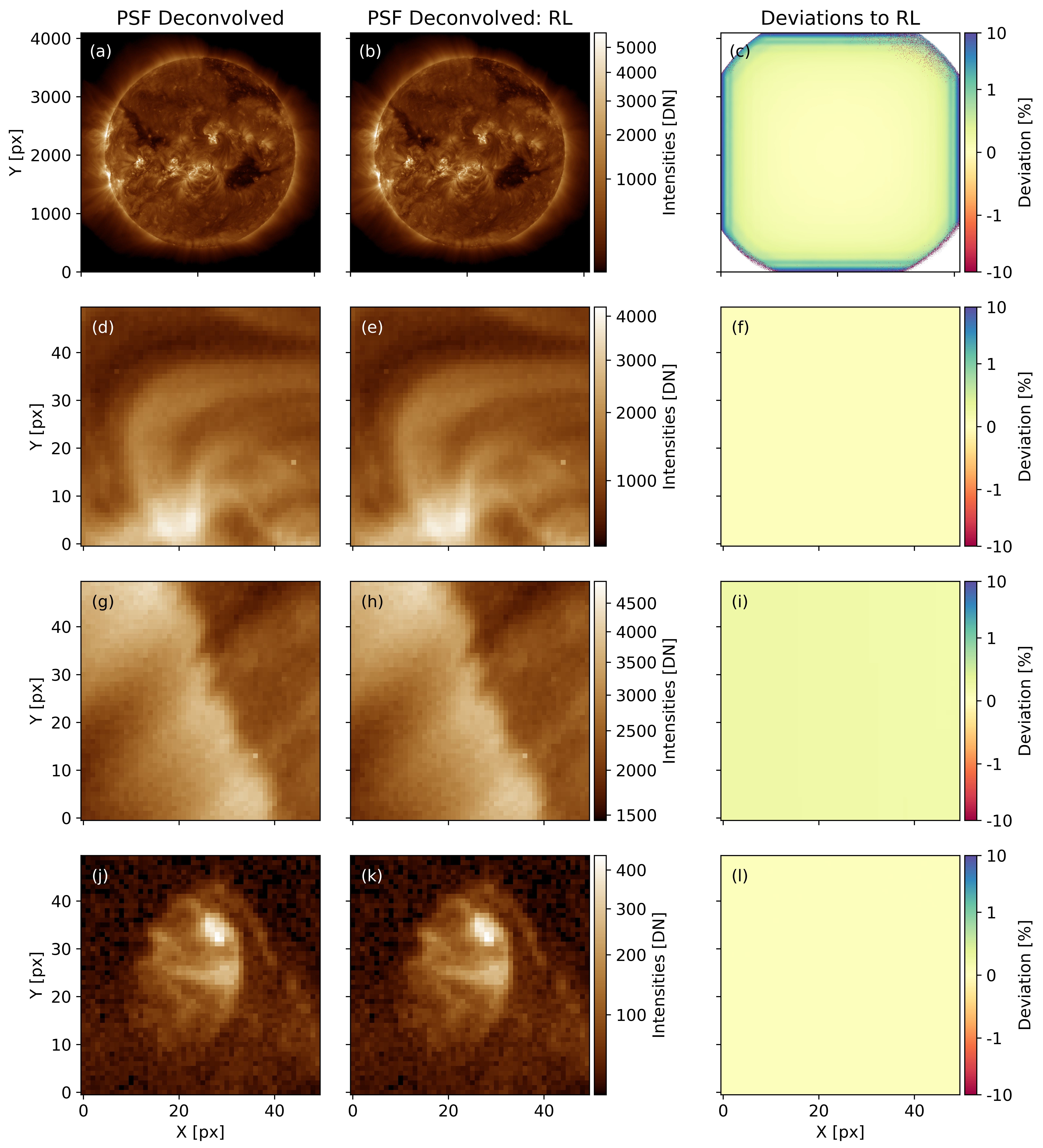}
    \caption{Comparison of the reconstruction results of BID and the Richardson-Lucy algorithm. The first row shows reconstruction results of the entire image of Fig.~\ref{fig:overview}. The  second to fourth row shows three zoom-ins into the regions marked in  Fig.~\ref{fig:overview} in green. First column: Reconstruction results by BID. Second column: Reconstruction results by the Richardson-Lucy algorithm. Third column: Percentage difference of the reconstructions of BID to the Richardson-Lucy algorithm.  }
    \label{fig:proof_of_concept}
\end{figure*}

\begin{figure*}
    \centering
    \includegraphics[width=\textwidth]{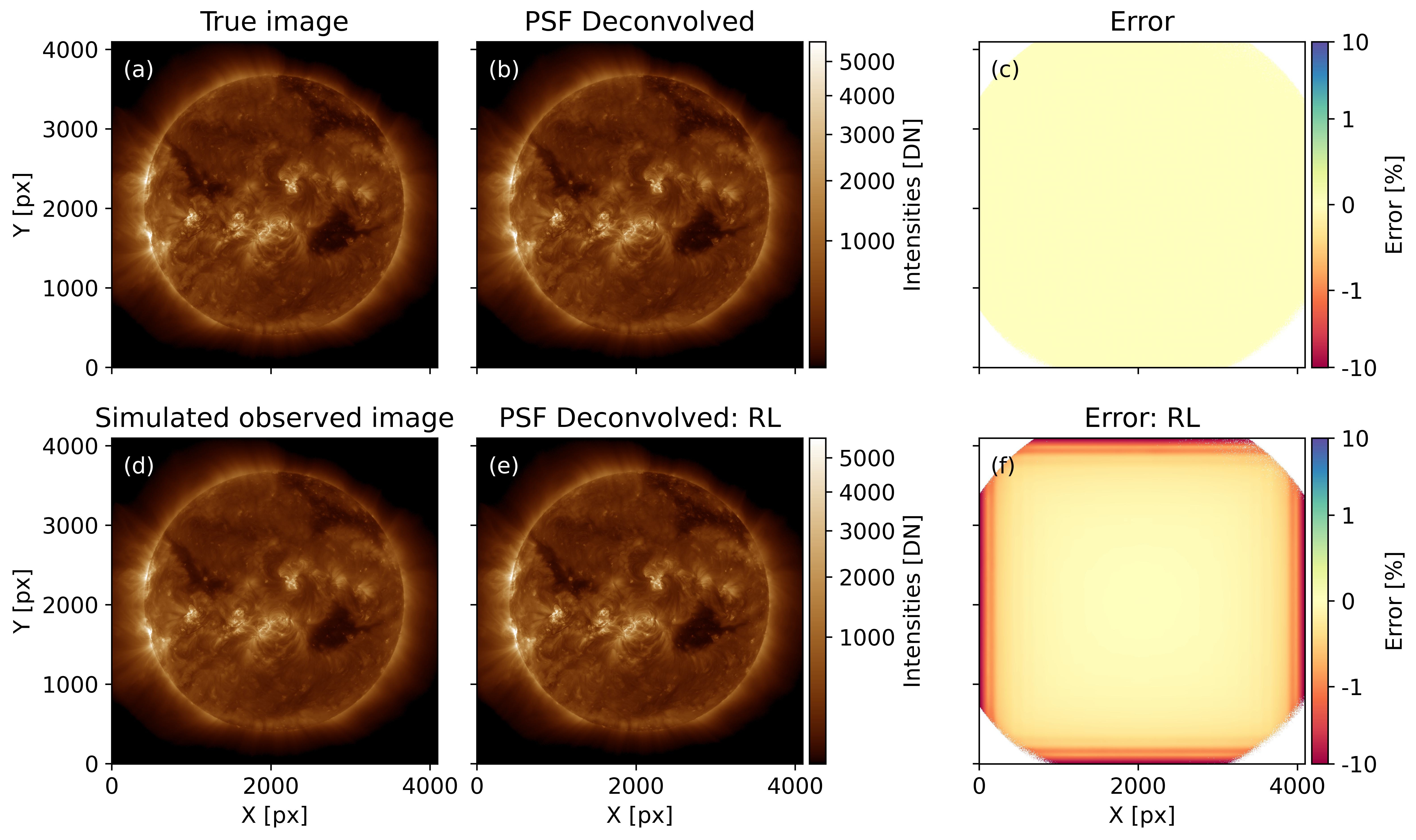}
    \caption{Artificial test. (a) Assumed true image. (b) Simulated observed image, derived by convolving the true image with the PSF. (c) and (d) Reconstruction results of BID and the Richardson-Lucy algorithm. (e) and (f) Percentage errors of the reconstructions as compared to the true image.  }
    \label{fig:proof_of_subimage1}
\end{figure*}

\begin{figure*}
    \centering
    \includegraphics[width=\textwidth]{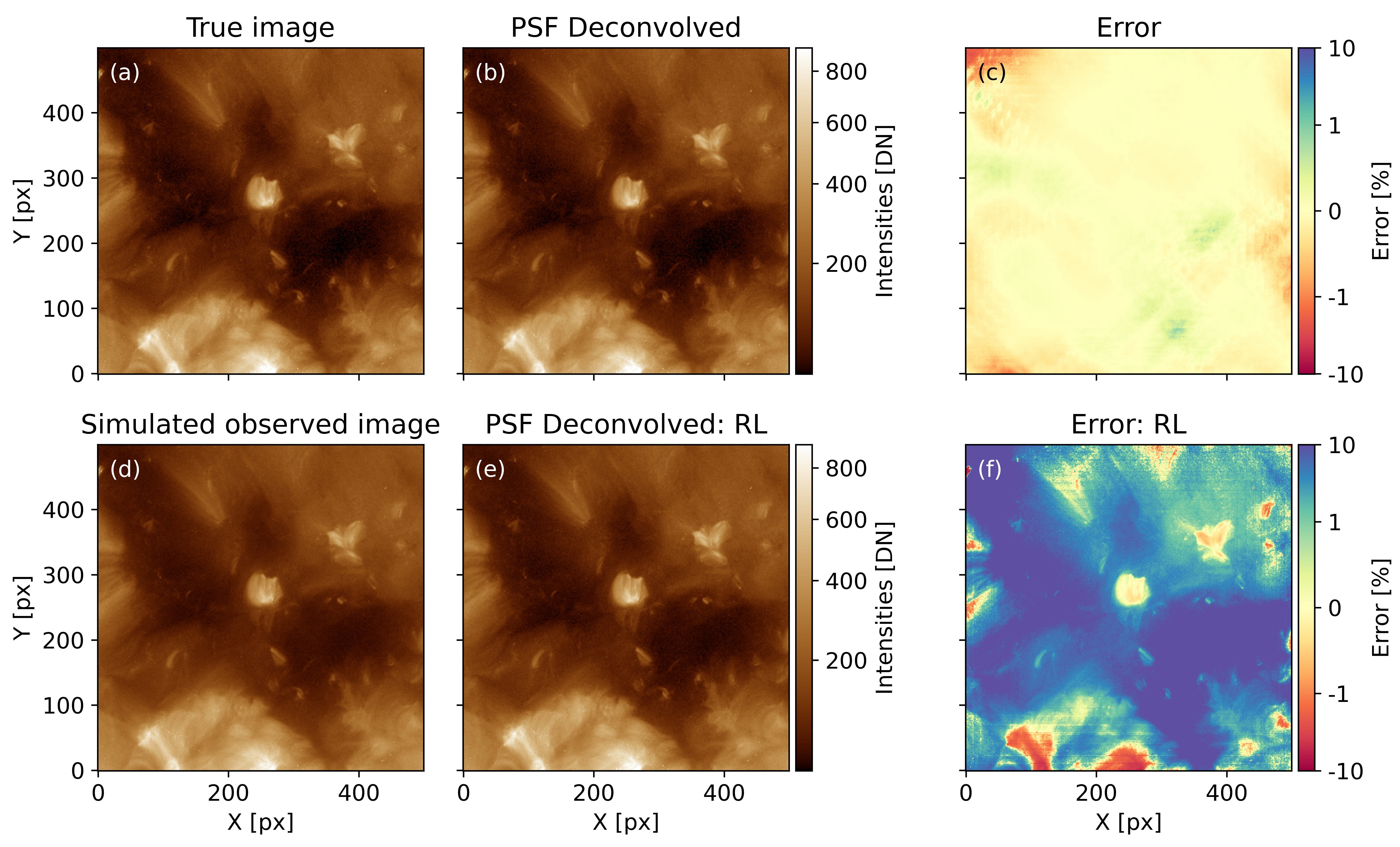}
    \caption{Same as Fig.~\ref{fig:proof_of_subimage1}, but for subimage reconstructions of the region marked in Fig.~\ref{fig:overview} in blue. }
    \label{fig:proof_of_subimage2}
\end{figure*}


We start with the fidelity of the reconstructions and subsequently proof that BID can retrieve photons that were scattered out of the image edges and that it can perform subimage deconvolutions. 

We consider BID reconstructions reliable if its reconstructions resemble the image reconstructions of the state-of-the-art Richardson-Lucy algorithm. To compare the reconstructions, for both algorithms, we first deconvolve the entire test image with the PSF. Then, we compare the reconstruction results  for the entire image and for the three cut-out subregions that are marked in Fig.~\ref{fig:overview} in green. The cutouts contain a loop system in an active region, a region containing the solar disk edge, and a coronal bright point in a coronal hole, and show the details regarding the similarities and differences between the reconstruction results. The reconstruction results are shown in Fig.~\ref{fig:proof_of_concept}. The entire test image reconstructions show that, for most parts, the reconstructions of BID and the Richardson-Lucy algorithm are in excellent agreement with differences \SI{<0.5}{\percent}. Only close to the image edges, differences of up to \SI{10}{\percent} are apparent. The reason for these differences will be explained in the next paragraph. The cutouts of the loop system, the coronal bright point, and the solar disk edge show that the reconstructions agree qualitatively and quantitatively remarkably well regarding all details up to the smallest spatial scales. 

Next, we simulate how well BID and the Richardson-Lucy algorithm can retrieve photons that are scattered out of the field of view of the detector. This is particularly important for solar extreme-ultraviolet imagers, as the PSFs of these imagers contain a strong diffraction pattern that scatters photons over large distances of several hundred pixels. 
For the simulation, we first assume the test image as true image. Then, we convolve the true image with the AIA PSF to derive an simulated observed image. Afterwards, we try to reconstruct the true image from the simulated observed image by deconvolving it with the AIA PSF. The error in the reconstructions as compared to the true image are shown in Fig.~\ref{fig:proof_of_subimage1}. For BID, the errors are everywhere smaller than \SI{0.01}{\percent}, i.e., BID excels in retrieving all photons that have been scattered out of the field of view of the detector. In contrast, the reconstructions by the Richardson-Lucy algorithm underestimates the intensities by up to \SI{10}{\percent} at the image edges and with declining errors towards the image center. The reason for this underestimation is that the Richardson-Lucy algorithm is flux-conserving within the image and thus cannot retrieve photons that are scattered out of the field of view of the detector.  

Finally, we simulate subimage deconvolutions. Again, we first assume the test image to be the entire true image and convolve it with the AIA PSF to derive the simulated observed image. Then, we cutout the in Fig.~\ref{fig:overview} in blue marked subregion and try to reconstruct its true intensities by deconvolving only the subregion with the AIA PSF. Finally, we analyze the errors in the reconstructions relative to the true intensities. We note that during the deconvolution, we use the extended BID algorithm with the incoming scattered light estimation. For the Richardson-Lucy algorithm, applying an incoming scattered light estimation is not sensible: the incoming scattered light from the surrounding into the subregion counteracts the loss of photons that are scattered out of the subregion; but since the Richardson-Lucy algorithm cannot retrieve photons that were lost by scattering out of the field of view, applying an incoming scattered light estimation alone worsens its reconstruction results. The errors of the subregion reconstructions relative to the true subregion intensities are shown in Fig.~\ref{fig:proof_of_subimage2}. For BID, the errors are \SI{<1}{\percent} in the entire subregion. In contrast, for the Richardson-Lucy algorithm, the reconstruction errors are \SI{>10}{\percent} for large parts of the subregion. These large errors are a consequence of its incapability to account for photons that were scattered out of and into the subregion. We conclude that, for the AIA PSF, the extended BID algorithm is suitable for subimage reconstructions while the Richardson-Lucy algorithm should not be used.



\subsection{Convergence and Speed} \label{sec:convergence}

\begin{figure*}
    \centering
    \includegraphics[width=\textwidth]{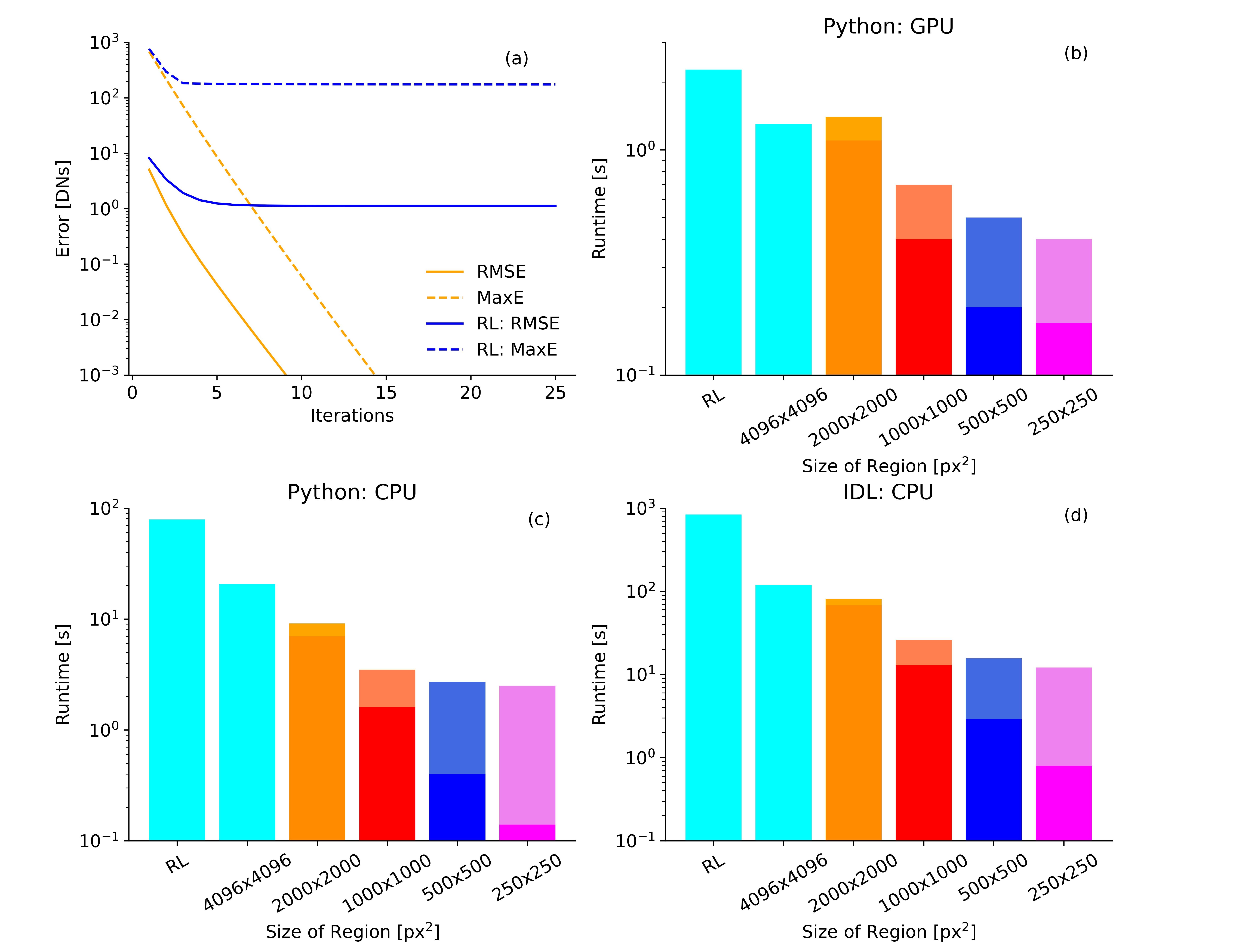}
    \caption{(a) Evolution of the RMSE and MaxE between the approximated true image convolved with the PSF and the observed image, for BID and the Richardson-Lucy (RL) algorithm. (b) Runtime of both algorithms in Python GPU implementations, (c) in Python CPU implementations, and (d) in IDL CPU implementations. The color of each bar corresponds to the corresponding marked subregions in Fig.~\ref{fig:overview}. In each bar associated with a subregion, the darker colors correspond to the runtime without applying the incoming scattered light estimation, and the lighter colors to the runtime with apllying the incoming scattered light estimation.}
    \label{fig:convergence}
\end{figure*}

We define the algorithm to have converged if the reconstructed approximation of the true image convolved with the PSF equals the observed image. Therefore, we quantify the quality of the convergence by the root mean square error (RMSE) and by the maximum error (MaxE) between the reconstructed true image convolved with the PSF and the observed image. In Fig.~\ref{fig:convergence}(a), we set both the BID algorithm and the Richardson-Lucy algorithm to run for $25$~iterations and plot the evolution of the RMSE and the MaxE versus the number of iterations. For BID, the RMSE and MaxE decrease steadily, reach a RMSE~of~\SI{0.0006}{DN} and a MaxE~of~\SI{0.1}{DN} at $10$~iterations, and continue to decrease until the limitation in the precision of our computing units is reached. In contrast, the RMSE and MaxE of the Richardson-Lucy algorithm decrease for $5$~iterations reaching a RMSE~of~\SI{1.5}{DN} and a MaxE~of~\SI{200}{DN}, and then stay at an almost constant level. 

There are two reasons why the convergence of the Richardson-Lucy algorithm saturates: (1) the Richardson-Lucy algorithm assumes photon noise which lets the algorithm converge statistically towards the true solution, while BID neglects photon noise which lets it solve the defining equation of PSF deconvolutions up to the machine precision. And (2), the Richardson-Lucy algorithm is flux conserving within the image and thus cannot retrieve photons that are scattered out of the detector, which results in a small residual reconstruction error. 

Next, we measure the runtime of both algorithms for a Python GPU, Python CPU, and an IDL CPU implementation. For BID, we measure the speed by performing the PSF deconvolution for the entire solar image image as well as for the four subregions of interest marked in Fig.~\ref{fig:overview} in orange, red, blue, and pink, which have sizes of $2000\times 2000$, $1000\times 1000$, $500\times 500$, and $250\times 250$~pixels. For the subregions, we perform the BID deconvolution one time with and one time without applying the incoming scattered light estimation. 
As for the Richardson-Lucy algorithm a subregion deconvolution is not sensible (cf.~Sect.~\ref{subsec:proof_of_concept}), we measure its speed  only for the entire image.
The runtimes are shown in Fig.~\ref{fig:convergence}(b)-(d). For the Python GPU implementations, the Richardson-Lucy algorithm runs for \SI{2.3}{seconds} for deconvolving the entire image, while BID needs \SI{1.3}{seconds} for the entire image, \SI{0.4}{seconds} for the $250\times 250$~pixels region of interest with the incoming scattered light estimation, and about \SI{0.2}{seconds} for the $250\times 250$~pixels region of interest without the incoming scattered light estimation. 
For the Python CPU implementations, the Richardson-Lucy algorithm runs for \SI{79}{seconds}, while BID needs \SI{21}{seconds} for the entire image, \SI{2.5}{seconds} for the $250\times 250$~pixels region of interest with the incoming scattered light estimation, and \SI{0.15}{seconds} for the $250\times 250$~pixels region of interest without the incoming scattered light estimation. 
For the IDL CPU implementations, the Richardson-Lucy algorithm runs for \SI{838}{seconds}, while BID needs \SI{119}{seconds} for the entire image, \SI{12.1}{seconds} for the $250\times 250$~pixels region of interest with the incoming scattered light estimation, and \SI{0.8}{seconds} for the $250\times 250$~pixels region of interest without the incoming scattered light estimation. 

\subsection{Noise Robustness}
\label{subsec:noiserobustness}

\begin{figure*}
    \centering
    \includegraphics[width=\textwidth]{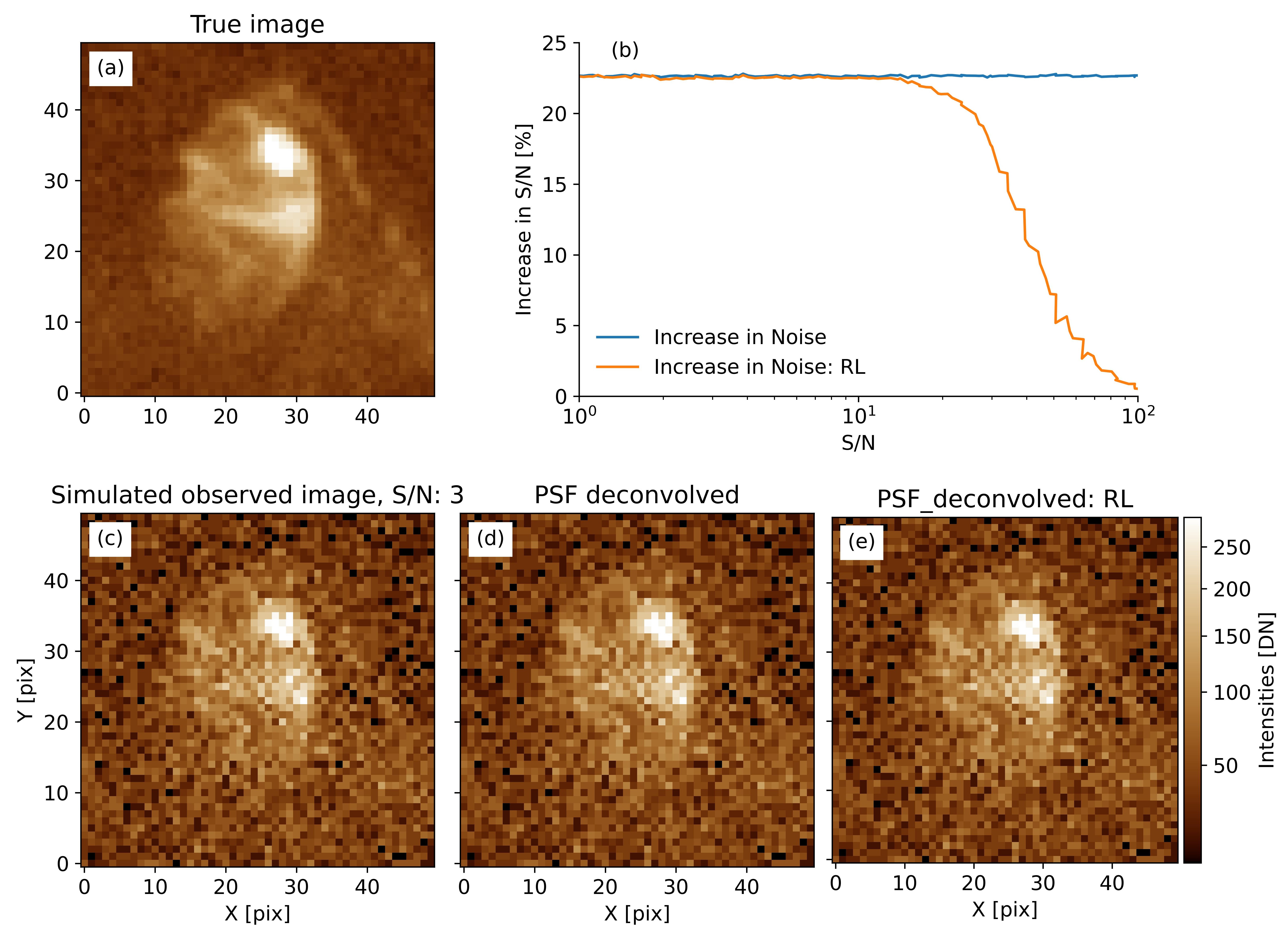}
    \caption{Simulations on the effect of image noise. (a) Assumed true image without noise. (b) Noise amplification as a function of the simulated signal-to-noise level for BID and the Richardson-Lucy algorithm. (c)-(e): Noisy simulated observed image and the reconstruction results for a signal-to-noise level of $3$.}
    \label{fig:noise}
\end{figure*}

As the BID algorithm does not constrain noise by regularization, it is in general less robust against image noise than the Richardson-Lucy algorithm. Here, we investigate if this is only a theoretical disadvantage or if it has a real impact on image deconvolutions with the AIA PSF.

 We  assume again the test image to be the true image and convolve it with the AIA PSF to derive the simulated observed image. Then, we vary the noise level in the observed image by simulating shorter exposure times, i.e., we re-draw the photon counts in each pixel by 
\begin{equation}
    I_\text{o,noise} = P(I_\text{o} / \text{noise} ) \cdot \text{noise},
\end{equation}
where $I_\text{o}$ is the observed intensity in a pixel without noise, $I_\text{o,noise}$ is the observed intensity in the pixel with added noise, $P$ is the Poisson distribution function, and $\text{noise}$ is the assumed noise level. Afterwards, we reconstruct the true images by deconvolving the noisy images with the PSF. Finally, we subtract from the noisy simulated observed image the simulated observed image without noise to obtain the original noise map and subtract from the reconstructed true image the true image to obtain the new noise map. We derive the noise levels in the original and new noise maps as the standard deviations of their intensities, and derive from that the noise amplification factor as the ratio of the noise levels. 

In Fig.~\ref{fig:noise}(b), we show the derived noise amplification factor as a function of the signal to noise level for the small subregion containing a coronal bright point. For the simulated signal-to-noise levels of 1--100, the BID algorithm constantly enhances the noise level by about \SI{23}{\percent}. The Richardson-Lucy algorithm enhances the noise level by about \SI{23}{\percent} for signal-to-noise levels of 1--10, and then improves to no noise enhancement at a signal-to-noise of $100$. In Fig.~\ref{fig:noise}(c)-(e), we show the noisy simulated observed image and the reconstruction results for a signal-to-noise level of $3$. The grainy structure arising from the noise in the reconstructed images from both algorithms resembles the grainy structure from the noise in the simulated observed image. Thus, the reconstructions only slightly amplify locally the existing noise, but they do not introduce new artifacts in the image. 

As expected, the Richardson-Lucy algorithm performs slightly better than the BID algorithm for noisy images. However, since for both algorithms the increase in noise is not large and well characterized, this increase in noise will not affect many analysis. If the region of interest in the observed image had a good signal-to-noise ratio before the reconstruction, it will still have a good signal-to-noise ratio after the reconstruction. If it had a bad signal-to-noise ratio before the reconstruction, one should anyway act with extreme caution when analyzing the data.

\subsection{Statistical analysis}

\begin{figure*}
    \centering
    \includegraphics[width=\textwidth]{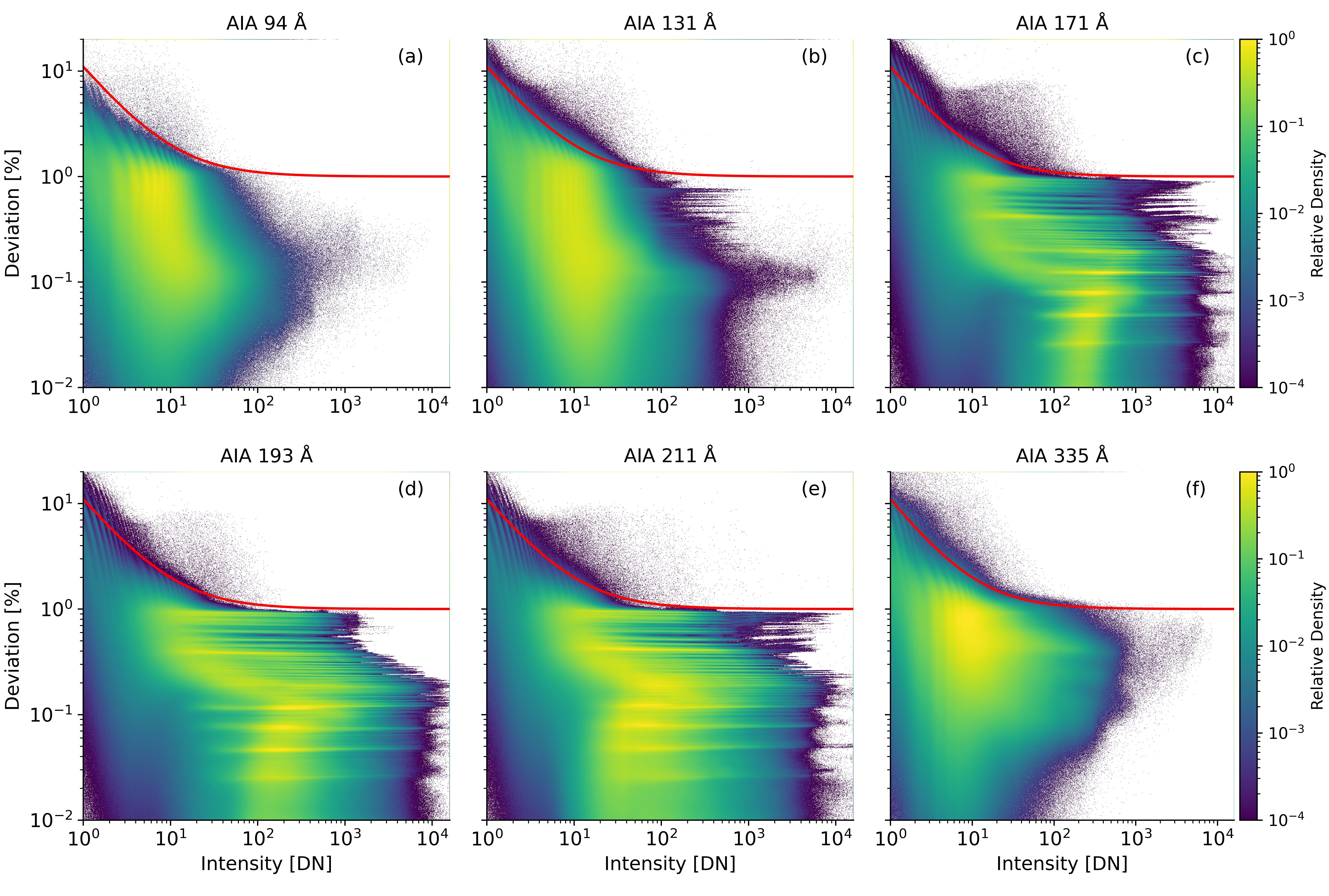}
    \caption{Density plots of the percentage deviations in the image reconstructions between BID and the Richardson-Lucy algorithm for AIA, dependent on the intensity level in the images. The red line corresponds to deviations of $\SI{1}{\percent} + \SI{0.1}{DN}$. }
    \label{fig:statistics}
\end{figure*}

Next, we compare the reconstruction results between BID  and the Richardson-Lucy algorithm in a large statistical dataset. For each of the six optical thin channels of AIA, the dataset comprises one image per month for the period October~2010 to February~2023. We first deconvolved the dataset with BID and with the Richardson-Lucy algorithm. We then selected only pixels where the Richardson-Lucy algorithm converged to better than \SI{1}{\percent} (cf.~Fig.~\ref{fig:convergence}(d)), and created from these pixels density plots of the percentage deviations between BID and the Richardson-Lucy algorithm, dependent on the image intensity levels. The results are presented in Fig.~\ref{fig:statistics}. In each panel, the density  has been normalized to the \num{99.93}~percentile of the  peak density in the panel. The red line marks a deviation in the reconstructed intensities of $\SI{1}{\percent} + \SI{0.1}{DN}$, where \SI{1}{\percent} is related to the quality of convergence of the selected pixels in the Richardson-Lucy algorithm and  \SI{0.1}{DN} corresponds to the uncertainty related to the BID termination condition. For more than \SI{97}{\percent} of the pixels in the AIA \SI{94}{\angstrom}, \SI{131}{\angstrom}, and \SI{335}{\angstrom} channels, and for more than \SI{99.5}{\percent} of the pixels in the \SI{171}{\angstrom}, \SI{193}{\angstrom}, and \SI{211}{\angstrom } channels, the deviations between the reconstructed intensities of BID and the ones from the Richardson-Lucy algorithm are smaller than \SI{1}{\percent} when the intensity level is larger than \SI{10}{DNs}, i.e., for regions with a reasonable signal-to-noise. This shows that the reconstruction results of BID and the Richardson-Lucy algorithm agree well as long as onne has a reasonable signal-to-noise level. Only at lower intensity levels, the percentage deviations between the reconstruction results increase. This is partially related to our chosen termination condition of \SI{0.1}{DN}, which gives an uncertainty of \SI{10}{\percent} at a \SI{1}{DN} intensity level. And it is partially to that image reconstructions of intensities close to the noise level are in general less certain, which affects the fidelity of the reconstruction results of the Richardson-Lucy algorithm, of BID, and thus also their concordance for these pixels.

\section{How to choose Reasonable Subregions } \label{sec:6}

\begin{figure*}
    \centering
    \includegraphics[width=\textwidth]{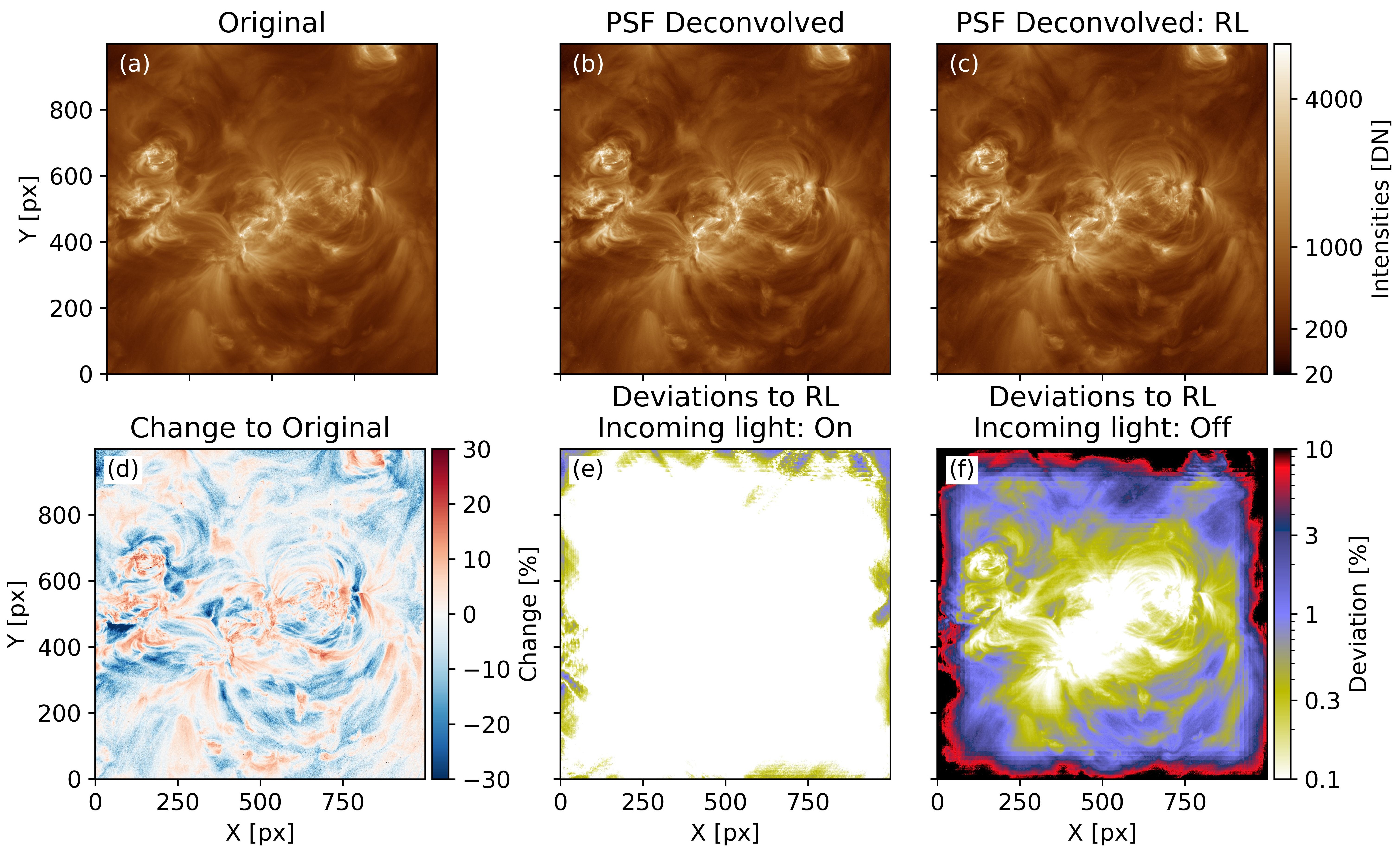}
      \caption{Subregion deconvolution of the active region marked in Fig.~\ref{fig:overview} in red. (a) Original image. (b) Reconstruction result of BID with the incoming scattered light estimation turned on. (c) Reconstruction results of the Richardson-Lucy algorithm. (d) Intensity change between the original image and our reconstructed image. (e) Percentage deviations between the reconstruction of BID when the incoming scattered light estimation is turned on and the Richardson-Lucy algorithm. (f): percentage deviations between the reconstruction of BID when the incoming scattered light estimation is turned off and the Richardson-Lucy algorithm.}
    \label{fig:example2}
\end{figure*}

\begin{figure*}
    \centering
    \includegraphics[width=\textwidth]{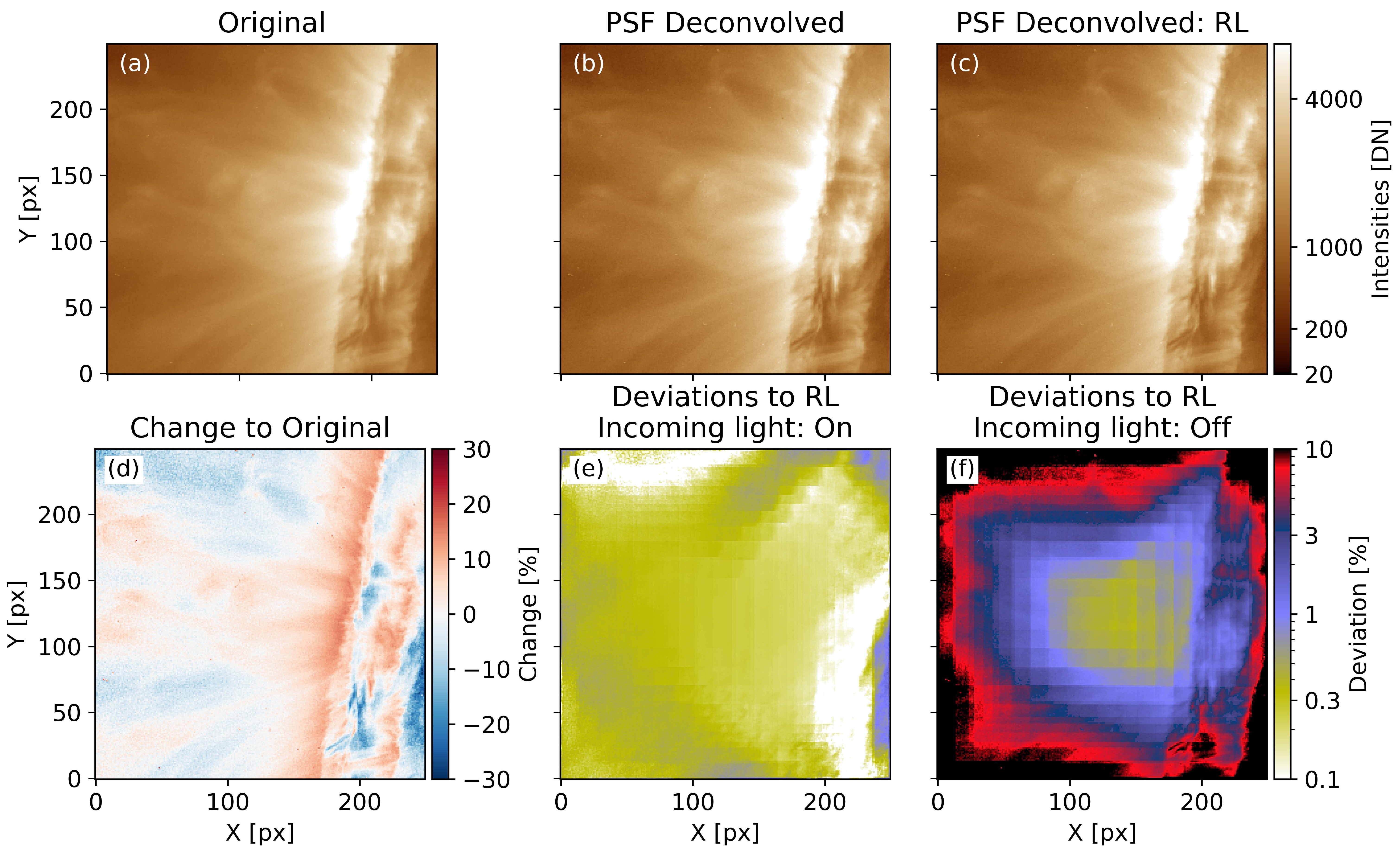}
    \caption{      
      Same as Fig.~\ref{fig:example2}, but showing the solar limb region marked in Fig.~\ref{fig:overview} in pink.}
    \label{fig:example1}
\end{figure*}

\begin{figure*}
    \centering
    \includegraphics[width=\textwidth]{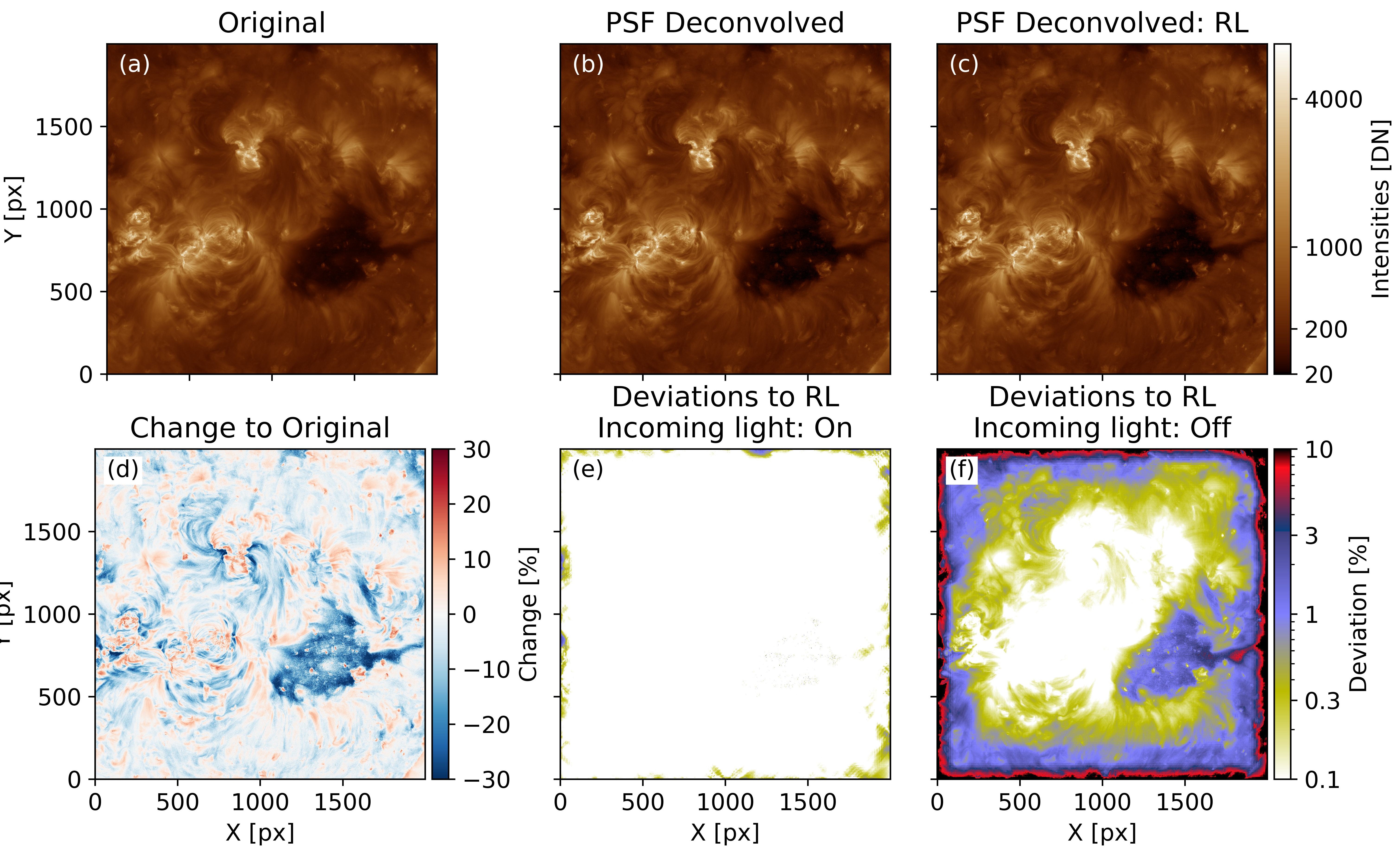}
       \caption{Same as Fig.~\ref{fig:example2}, but showing the coronal hole region marked in Fig.~\ref{fig:overview} in orange.}
    \label{fig:example3}
\end{figure*}

Here, we will shortly demonstrate how to choose reasonable subregions for subimage deconvolutions. We will show this on the example of three subregions, i.e., an active region (marked in Fig.~\ref{fig:overview} in red), a quiet Sun region at the solar limb (pink), and a coronal hole (blue). 

The following figures show in panel (a) the observed subregion, (b) the deconvolved subregion using a BID subregion deconvolution, (c) the cutout subregion from a full-image Richardson-Lucy deconvolution, and (d) the intensity change from the observed subregion to the BID reconstructed subregion. For these four panels, the incoming scattered light estimation for BID was turned on. Panel (e) shows the deviations between the reconstruction results between BID and the Richardson-Lucy algorithm with the incoming scattered light estimation for BID turned on, and panel (f) shows the deviations with the incoming scattered light estimation being turned off.

Figure~\ref{fig:example2} targets an active region. As active regions are typically \numrange{1}{2} orders of magnitude brighter than the quiet Sun, the subregion can be chosen to tightly enclose the target active region. The incoming scattered light from the surrounding to the active region can be neglected. Panels (e) and (f) show the deviations of the reconstruction results to the Richardson-Lucy algorithm with the incoming scattered light estimation for BID being turned on (e) and off (f). For both cases, the agreement between the active region reconstruction results of the BID subregion deconvolution and the full-image Richardson-Lucy deconvolution is better than \SI{0.3}{\percent}. Only if the interest lies in the surrounding of the active region, the incoming scattered light estimation is required to be turned on.

Figure~\ref{fig:example1} targets a quiet Sun loop system at the solar disk edge. The loop system has a similar intensity as its surrounding, therefore, the incoming scattered light can be important and the incoming scattered light estimation should be turned on. As there are no bright active regions in its near vicinity, the subregion can be chosen to rather tightly enclose the target loop system (however, if there would be active regions nearby, they should be included in the subregion). Panels (e) and (f) show that only the BID subregion deconvolution with the incoming scattered light estimation turned on gives satisfactory results with deviations to the full-image Richardson-Lucy deconvolution being \SI{<0.5}{\percent}. To further increase the accuracy, one could increase the size of the selected subregion.

Finally, Figure~\ref{fig:example3} targets a coronal hole, which is \numrange{1}{2} orders of magnitude darker than the surrounding quiet Sun and \numrange{2}{3} orders of magnitudes darker than active regions. Therefore, scattered light from the surrounding into the coronal hole has a significant effect on the coronal hole intensities, and consequentially, the incoming scattered light estimation must be turned on. Furthermore, one has to select a generously large subregion which contains the coronal hole, the surrounding quiet Sun, and all nearby active regions. With these settings, Panel (e) shows that the subregion reconstruction is accurate with deviations to the Richardson-Lucy algorithm being \SI{<0.2}{\percent}.

Summarizing, the subregion should be selected to contain all nearby bright regions to the target; and in case that the target is not a very bright region itself, the incoming scattered light estimation should be turned on; and in case that the target is dark, the size of the subregion should be selected to be generously large.

\section{Limitations and Discussion} \label{sec:5}

\begin{figure*}
    \centering
    \includegraphics[width=\textwidth]{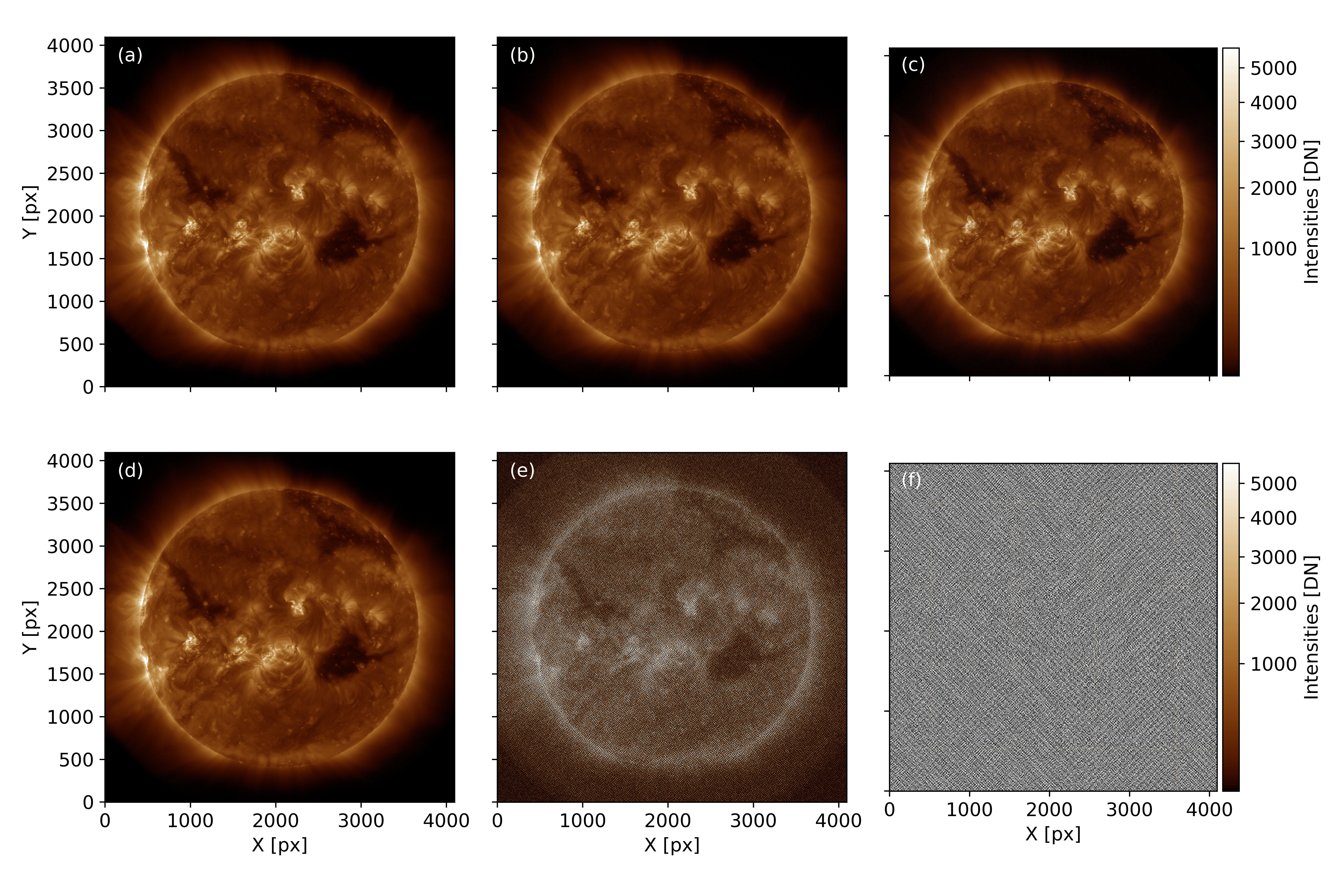}
    \caption{      
      Stability of the inversion algorithm. The first row shows the PSF deconvolved image using BID, and the second row the PSF deconvolved image using the Inverse Filter. In first column, the PSF is a Gaussian with a sigma of~\SI{1}{px}, in the second column a Gaussian with a sigma of~\SI{1.3}{px}, and in the third column a Gaussian with a sigma of \SI{2}{px}.}
    \label{fig:comparison_inv}
\end{figure*}

The results presented targeted specifically the solar community, and, as such, the instruments and PSFs they use. Here, we want to set the BID algorithm and the option of subimage deconvolutions in a larger context. 

In the image deconvolution community, it is usually assumed that deconvolution algorithms which do not constrain the noise during the deconvolution amplify the noise to unacceptable levels. From a mathematical perspective, which focuses on developing general deconvolution algorithms, this is true. However, there are subfields in the space of image deconvolution use cases where the noise only becomes mildly amplified. One of them is the use case of image deconvolutions with the instrumental PSFs of well-focused imagers. According to Sect.~\ref{subsec:noise}, for iterative image deconvolutions in the spatial domain with the PSF of well-focused imagers, the noise amplification factor depends mostly on the center coefficient of the PSF, i.e., on the amount of light that is not scattered in the instrument. In these cases, the center coefficient is sufficiently large so that the image noise only becomes mildly amplified. Thus, for instrumental PSFs, such as for the AIA PSF, image reconstructions with the BID algorithm are robust. However, for non-instrumental PSF, such as PSFs that describe seeing conditions and motion blurring, the center coefficient is typically very small and the importance of the surrounding PSF core coefficients increases. In these cases, the noise amplification can become large and other deconvolution algorithms which constrain the noise are to be preferred. 

BID is mathematically equivalent to the Inverse Filter. Both solve the PSF defining equation, Eq.~\ref{eq:psf} without a regularization and neglect the noise component. While BID operates in the spatial domain, the Inverse Filter performs the deconvolution by a division in the frequency domain, 
\begin{align}
     \mathcal{F}(\mathbf{I}_\text{t}) &= \frac{\mathcal{F}(\mathbf{I}_\text{o}) - \mathcal{F}(\bm{\mathcal{E}})}{\mathcal{F}(\mathbf{PSF})}, \label{eq:invfilt}
\end{align}
where~$\mathcal{F}$ denotes the Fourier transformation. 
Figure~\ref{fig:comparison_inv} shows the deconvolution results of BID and the Inverse Filter for PSFs consisting of a Gaussian with a sigma of \SI{1}{px}, \SI{1.3}{px}, and \SI{2}{px}, i.e., where the PSF center coefficients are~\SI{16}{\percent},~\SI{9}{\percent}, and~\SI{4}{\percent}. While our implementation of BID stays stable for all these PSFs, the Inverse Filter becomes unstable and amplifies noise to unacceptable levels for the PSFs with Gaussian sigmas of~\SI{1.3}{px} and~\SI{2}{px}. The reason is found in the iterative evolution of BID's deblurring versus noise amplification characteristics. In the first iterations, BID robustly deblurs images and only slightly amplifies the noise. Once BID reaches a critical point, it would start to amplify noise and, after an infinite number of iterations, slowly converge to the Inverse Filter solution. By terminating BID when an acceptable accuracy in the deconvolution is reached, i.e., \SI{0.1}{DN} in our implementation, one keep BIDs robust deblurring characteristics and avoid the excessive noise amplification of the Inverse Filter. For a more detailed discussion of this topic, we recommend the review paper on iterative deconvolution algorithms of \citet{biemond1990}.

The feasibility of subimage deconvolutions for different kinds of deconvolution algorithms depends on the PSFs used. For many instrumental PSFs, the PSF can be approximated as a gaussian core with rather short wings, i.e., light is scattered only over short distances. In this case, the incoming scattered light from the surrounding into the subregion and the light scattered out of the subregion can be neglected, and consequentially, almost every deconvolution algorithm can be used for these subregion deconvolutions. However, when the PSF has a component that scatters light over large distances, such as the diffraction patterns in the AIA PSFs, the incoming light and the light scattered out of the subimage have to be treated properly. The incoming light can be statistically estimated by the preprocessing step described in Sect.~\ref{sec:2} for all algorithms. However, the amount of light scattered out of the subimage depends on the solution for the true subimage and, as such, has to be estimated during the deconvolution process itself. The exact solution for the light scattered out of the subimage is given by the defining equation of PSF deconvolutions itself (cf. Sect.~\ref{sec:conservationofflux}). As the BID algorithm simply solves the defining equation of the PSF without any further constraints (besides neglecting statistical image noise), it excels in estimating the amount of light that is scattered out of the subimage. For this reason, BID is especially well suited for subimage deconvolutions when the PSF scatters light over large distances, such as the AIA PSF in solar physics.

\section{Summary} \label{sec:4}
We presented the BID algorithm, which we extended for subregion image deconvolutions, and compared the image reconstruction results to those from the Richardson-Lucy algorithm. The presented algorithm
\begin{itemize}
    \item is faster than the Richardson-Lucy algorithm by a factor of $1.8$  when comparing Python GPU implementations, by a factor of $3.7$ when comparing Python CPU implementation, and by a factor of $7.1$ when comparing IDL CPU implementations,
    \item is able to deconvolve image subregions, increasing its speed by an additional factor of up to $150$ for subregions having a size of $250\times 250$~pixels, 
    \item can consider light scattered from the surrounding image into the subregion of interest,
    \item accounts for photons that are scattered out of the subregion of interest, or, respectively, the detector,
    \item increases the noise level slightly more than the Richardson-Lucy algorithm for instrumental PSFs,
    \item but is still accurate, with typical deviations of less than $\SI{1}{\percent}$ to the reconstruction results of the Richardson-Lucy algorithm for AIA images.
\end{itemize}
This algorithm is especially well suited to correct satellite-born solar  images for instrumental effects. It can retrieve photons that were scattered out of the field of view of the detector, which is typically the case for solar images taken in the extreme ultraviolet. And it is significantly faster than the Richardson-Lucy algorithm for subregion reconstructions, which enables one to reconstruct long sequences of images, such as for studies on the evolution of solar features, in a limited period of computation time. Due to these advantages, the extended BID algorithm can replace the Richardson-Lucy algorithm for a variety of use cases in solar physics.

 \begin{acknowledgements}
I thank Daniel W. Savin for carefully reading the manuscript and for his outstanding textual remarks. I acknowledge support from the German Science Fund (DFG) under project number 448336908. The AIA images are provided by courtesy of NASA/SDO and the AIA, EVE, and HMI science teams.

 \end{acknowledgements}

\bibliographystyle{aa} 
\bibliography{bibliography}

\end{document}